\documentclass[prd,12pt,amssymb,amsfonts,amsmath]{revtex4}
\usepackage{epsfig}

\newcommand{\Det}{\mathrm{Det }\,\,}
\newcommand{\reals}{\mathbb{R}}

\begin{document}
\title{Area metric gravity and accelerating cosmology}

\author{Raffaele Punzi}
  \affiliation{Dipartimento di Fisica ``E. R. Caianiello'' Universit\`a di Salerno, 84081 Baronissi (SA) Italy}
  \affiliation{INFN - Gruppo Collegato di Salerno, Italy}
\author{Frederic P. Schuller}\affiliation{Instituto de Ciencias Nucleares, Universidad Nacional Aut\'onoma de M\'exico, A. Postal 70-543, M\'exico D.F. 04510, M\'exico}
\author{Mattias N. R. Wohlfarth}
\affiliation{Center for Mathematical Physics and {II}. Institut f\"ur Theoretische Physik, Luruper Chaussee 149, 22761 Hamburg, Germany}
\begin{abstract}
Area metric manifolds emerge as effective classical backgrounds in quantum string theory and quantum gauge theory, and present a true generalization of metric geometry. Here, we consider area metric manifolds in their own right, and develop in detail the foundations of area metric differential geometry. Based on the construction of an area metric curvature scalar, which reduces in the metric-induced case to the Ricci scalar, we re-interpret the Einstein-Hilbert action as dynamics for an area metric spacetime. In contrast to modifications of general relativity based on metric geometry, no continuous deformation scale needs to be introduced; the extension to area geometry is purely structural and thus rigid. We present an intriguing prediction of area metric gravity: without dark energy or fine-tuning, the late universe exhibits a small acceleration.
\end{abstract}
\maketitle
\newpage
\section*{INVITATION}
A new theoretical concept which, once formulated, naturally emerges
 in many related contexts, deserves further study. Even more so,
if it makes us view
well-established theories in a novel way, and meaningfully points
beyond standard theory.

Area metrics, we argue in this paper, are such an emerging
notion in fundamental physics. An area metric may be defined as a
fourth rank tensor field which allows to assign a measure to
two-dimensional tangent areas, in close analogy to the way a
metric assigns a measure to tangent vectors. In more than three
dimensions, area metric geometry is a true generalization of
metric geometry; although every metric induces an area metric,
not every area metric comes from an underlying metric. The
mathematical constructions, and physical conclusions, of the
present paper are then based on a single principle:
\begin{center}
  \emph{Spacetime is an area metric manifold.}
\end{center}
We will be concerned with justifying this rather bold idea by a
detailed construction of the geometry of area metric manifolds,
followed by providing an appropriate theory of gravity, which
finally culminates in an application of our ideas to cosmology. In
the highly symmetric cosmological area metric spacetimes, we can
compare our results easily to those of Einstein gravity. We obtain
the interesting result that the simplest type of area metric
cosmology, namely a universe filled with non-interacting string
matter, may be solved exactly and is able to explain the
observed~\cite{Spergel:2003cb,Knop:2003iy} very small late-time
acceleration of our Universe, see the figure on page \pageref{figure1},
without introducing any notion of dark energy, nor by invoking
fine-tuning arguments.

It may come as a surprise, but standard physical theory itself
predicts the departure from metric to true area metric manifolds.
More precisely, the quantization of classical theories based on
metric geometry generates, in a number of interesting cases, area
metric geometries: back-reacting
photons in quantum electrodynamics effectively propagate in an
area metric background \cite{Drummond:1979pp}; the massless states of
quantum string theory give rise to the Neveu-Schwarz two-form
potential and dilaton besides the graviton, producing a
generalized geometry which may be neatly absorbed into an area metric
\cite{Schuller:2005ru}; the low energy action for D-branes
\cite{Fradkin:1985qd,Abouelsaood:1986gd,Bergshoeff:1987at,Leigh:1989jq}
is a true area metric volume integral \cite{Schuller:2005ru};
canonical quantization of gravity \`a la
Ashtekar~\cite{Ashtekar:1986yd,Rovelli:1989za}
naturally leads to an area operator~\cite{Rovelli:1994ge},
such that the classical limit
of the underlying spin network structure is also likely a generic
area metric manifold, rather than a metric one.

The emerging picture is that area metric manifolds are generalized
geometries. In the case of string theory, one may even reverse the
argument by observing that the geometry of an area metric
background forces one to
consider strings rather than point particles~\cite{Schuller:2005yt}.
In the present paper this
plays a role in our discussion of fluids in area cosmology;
fluids on area metric spacetime cannot consist of particles, but
must feature strings as the minimal mechanical objects, which leads
us to develop the notion of a string fluid.

Generalized geometries begin to play an
increasingly important role also in mainstream string theory, despite the
fact that one initial starting point for its formulation is a
metric target space manifold. Non-geometric backgrounds in string
theory, meaning backgrounds that do not admit a metric geometry,
for instance emerge in flux compactifications on which one acts
with T-dualities or in mirror symmetry
\cite{Kachru:2002sk,Hellerman:2002ax,Gurrieri:2002wz,Fidanza:2003zi}. They also
appear in compactifications with duality twists, which in
some cases have been shown to be equivalent to asymmetric
orbifolds~\cite{Dabholkar:2002sy,Hull:2003kr,Flournoy:2004vn}.
Generalized geometries built to understand
these situations have been originally proposed by Hitchin
\cite{Hitchin:2004ut,Gualtieri:2004} and, the
T-fold idea, by Hull \cite{Hull:2004in,Dabholkar:2005ve}.
These have found a number of
applications~\cite{Grana:2004bg,Grana:2005ny,
Koerber:2005qi,Zucchini:2005rh,Zabzine:2006uz,Reid-Edwards:2006vu,Grange:2006es},
one recent example discusses the
stabilization of all moduli through fluxes in a specific
non-geometric background~\cite{Becker:2006ks}.

Maybe the most striking example of a classical theory, where area
metrics play a natural role, is gauge theory in general, and
Maxwell electrodynamics in particular. Given that electrodynamics is the
historical birth place of the concept
of a spacetime metric, this is certainly noteworthy. It is known that
electrodynamics may be formulated on any $d$-dimensional smooth manifold
without even introducing the concept of a metric
\cite{Lammerzahl:2004ww,Hehl:2004yk,Hehl:2005xu}. The
idea of this so-called pre-metric approach goes back to a paper by
Peres \cite{Peres:1962}, and is based on the observation that charges, and
in certain situations, magnetic flux lines can be counted, which
is nicely explained in \cite{Hehl:2004yk}. Hence one may define the notions
of a field strength two-form $F$ and an electromagnetic induction
$(d-2)$-form $H$. The equations of vacuum electrodynamics are then
given by $dF=0$ and $dH=0$. The induction
two-tensor~$\mathcal{H}$ dual to~$H$ must
be related to the field strength $F$ by some constitutive
relation in order to close the system of
equations. While Peres originally took this
to be a definition of the metric, Hehl et al. \cite{Hehl:2004yk}
generalized this to an arbitrary linear relation $\mathcal{H} = \chi F$
described by a tensor $\chi$ of
fourth rank, and investigated which conditions imply light
propagation along a Lorentzian lightcone. Such general relations
between the field strength and the induction are known from the description of
electrodynamics in continuous media; the lensing of light
rays in some materials cannot be described by
geodesics in a metric background. It is therefore not too daring
to suspect that gravitational lensing may be equally rich, which
amounts to the assumption that spacetime is an area metric
manifold---our central assumption in this paper.

Of course, neither the most extensive list of known phenomena, nor the
most suggestive hints for generalizations alone will be able to
justify our proposal to consider spacetime an area metric
manifold. However, the area metric gravity theory developed in this
paper seems a particularly worthwile testbed for our hypothesis;
this is not least due to the fact that we do not introduce any new
parameter into the theory. In this sense, it is not a deformation
of Einstein-Hilbert gravity (as every alternative action based on
metric geometry necessarily is), but rather an extension in which
the metric Ricci scalar is replaced by its area metric analogue.
The prediction of an accelerated expansion of our Universe at late
times, without any additional assumptions or scales being put in
by hand, should count as a promising indication in favour of the
idea of area metric spacetime.

\hspace{12pt}

This paper is divided into two largely self-contained parts. Part One (sections 1--7) develops the foundations of area metric geometry
in detail. Its practical results, however, are concisely summarized in the first section of Part Two (sections 8-14) where the area metric version of Einstein-Hilbert gravity is formulated in general, and then applied to area metric cosmology. A more detailed outline of the individual sections of the paper is given at the beginning of each of the two parts.
We conclude the paper with a discussion of our results and point out future directions. Appendix~\ref{conventions} lists our conventions, while appendices~\ref{appvary} and~\ref{Qsimplification} respectively derive the general equations of motion of four-dimensional area metric gravity, and those simplified for the almost metric case.

\newpage
\section*{P A R T \quad O N E :\\ A R E A \quad  M E T R I C \quad  G E O M E T R Y}
This first purely mathematical part of the paper may be skipped at first reading. Its practically relevant results are concisely summarized at the beginning of Part Two, so that the reader mainly interested in the application of area metric geometry to gravity and cosmology may fast-forward to the second part, and come back later to the in-depth treatment of area metric geometry presented here.

The precise definition of area metrics, densities and volume forms is given in section~\ref{manifolds}, before the non-linear structure of the space of oriented areas is briefly discussed in section~\ref{sec_areabundles}. The ensuing construction of area metric geometry is canonical in the sense that it does not rely on additional structure beyond area metric data \cite{foot1}. A central issue, namely the extraction of some effective metric from an area metric, is resolved in three steps: identification of the Fresnel tensor associated with an area metric in section~\ref{gauge}, construction of a family of pre-metrics from the Fresnel tensor in section~\ref{normal}, and finally selection of a unique, non-degenerate member of that family in section~\ref{sec_effmetric}. The aside on area metric symmetries in section~\ref{sec_Killing} prepares our discussion of cosmology later on. From a practical point of view, the most important result of this first part of the paper is the construction of area metric curvature tensors which are downward compatible to their metric counterparts, in section~\ref{curvature}.

\section{Area metric manifolds}\label{manifolds}
Knowing how to measure lengths and angles, one knows how to measure areas. More precisely, if $(M,g)$ is a metric manifold, one may define the tensor
\begin{equation}\label{Cg}
  C_g(X,Y,A,B) = g(X,A)g(Y,B) - g(X,B)g(Y,A)
\end{equation}
that measures the squared area of a parallelogram spanned by vectors $(X,Y)$ as $C_g(X,Y,X,Y)$. We will call $C_g$ the area metric induced from the metric $g$.

The basic idea of area metric geometry consists in promoting area metrics to a structure in their own right, independent of whether there is some underlying metric or not. To achieve this generalization, we simply introduce the area metric by keeping some of the salient algebraic properties of the metric-induced area metrics (\ref{Cg}). Formally, we define an area metric manifold $(M,G)$ as a smooth $d$-dimensional manifold $M$ equipped with a fourth-rank covariant tensor field $G$ that satisfies the following symmetry and invertibility properties at each point of the manifold:
\begin{enumerate}
     \item[{\sl (i)}] $G(X,Y,A,B) = G(A,B,X,Y)$ for all vector fields $X,Y,A,B$ in $TM$,
     \item[{\sl (ii)}] $G(X,Y,A,B) = - G(Y,X,A,B)$ for all vector fields $X,Y,A,B$ in $TM$,
     \item[{\sl (iii)}] $G: \Lambda^2TM \to \Lambda^2T^*M$, defined as $G(\Omega)(\Sigma) = G(\Omega,\Sigma)$ by continuation, is invertible.
\end{enumerate}
Here $\Lambda^2T_pM$ denotes the space of all contravariant
antisymmetric tensors of rank two; for our conventions concerning
components of $\Lambda^2TM$ tensors see appendix
\ref{conventions}. We will see at the end of section \ref{gauge}
that in three dimensions every area metric is metric-induced; from
four dimensions onwards, however, there exist area metrics that
cannot be induced from any metric.

In addition to the symmetries {\sl (i)} and {\sl (ii)} which we included in our definition, a metric-induced area metric (\ref{Cg}) features a third symmetry,
namely cyclicity:
\begin{equation}
     C(A,X,Y,Z) + C(A,Y,Z,X) + C(A,Z,X,Y) = 0 \quad \textrm{for all $A,X,Y,Z$ in $TM$}\,.
\end{equation}
We emphasize that we do not impose cyclicity as a property of generic area metrics. In fact, we will see shortly
that non-cyclicity plays a central role in two related problems: in the extraction of metric information from
an area metric and in the construction of area metric compatible connections.
The price to pay for non-cylicity is that the area metric tensor $G$ is algebraically reducible. More precisely, any
area metric decomposes uniquely into a cyclic area metric and a four-form, which are both irreducible. It will
turn out to be advantageous for technical reasons to consider such a decomposition for the inverse area metric,
\begin{equation}\label{C+F}
  G^{abcd} = C^{abcd} + F^{[abcd]}\,.
\end{equation}
In this context
note that the cyclic components $C$ and the four-form components $F$ of the inverse area metric generically mix under inversion;
in other words, the cyclic part of the inverse area metric is not simply
the inverse of the cyclic part of the area metric.

An area metric $G$ naturally gives rise to the scalar density
$|\Det G|^{1/(2d-2)}$ of weight $+1$, where the
capitalized determinant $\Det\!$ is understood to be taken over the square matrix of dimension $d(d-1)/2$ representing the map $G: \Lambda^2TM \times \Lambda^2TM \to \mathbb{R}$.
The correct transformation behaviour under diffeomorphisms is easily seen by direct calculation, noting that for any square matrix $T$
of dimension $d$ we have
\begin{equation}
   \Det T^{[a_1}{}_{b_1} T^{a_2]}{}_{b_2} = (\det T^a{}_b)^{d-1}\,,
\end{equation}
where the determinant $\det$ denotes the standard determinant. Hence any area metric manifold is naturally equipped with a volume form $\omega_G$, whose components in some basis are given by
\begin{equation}
     \omega_{G\,\,}{}_{a_1 \dots a_d} = |\mathrm{Det } G|^{1/(2d-2)} \epsilon_{a_1 \dots a_d}\,,
\end{equation}
where $\epsilon$ is the Levi-Civita tensor density normalized such that $\epsilon_{1 \dots d}=1$.

Now that we have given a precise definition of area metric manifolds, we should mention some relations to the mathematical literature. The idea to base geometry on some measure of area goes back to work by Cartan \cite{Cartan:1933}, and has been generalized under the name of `areal spaces', see \cite{Barthel:1959, Brickell:1968,Davies:1972,Davies:1973,Tamassy:1995} and references therein, to geometries based on reparametrization-invariant integrals, i.e., to any given volume measure. Although these geometries are more general than our physically motivated area measure, this does not mean that we will simply reproduce, or specialize, known mathematics in this paper. In fact, it is precisely our tensorial approach that allows us, in a novel way, to construct an effective metric and connections on the embedding bundle of areas. Moreover the physical motivation behind our construction leads to a successful application to gravity theory.

\section{Area bundles}\label{sec_areabundles}
At first sight, it seems that an area metric assigns a measure to
parallelograms. But actually, an area metric $G$ assigns equal
area measure to any two co-planar parallelograms $(X,Y)$ and
$(\tilde X, \tilde Y)$ in $T_pM \oplus T_pM$ that are related by an
$SL(2, \reals)$ basis transformation. To see this, let $\tilde X = aX+
bY$ and $\tilde Y = cX + dY$ with $ad-bc=1$; then
\begin{equation}
  G(\tilde X, \tilde Y, \tilde X, \tilde Y) = (ad-bc)^2 G(X,Y,X,Y)\,.
\end{equation}
The equivalence class of all parallelograms that are $SL(2,\reals)$-related
to some representative parallelogram $(X,Y)$ is algebraically neatly realized as the wedge product $X \wedge Y$, since
\begin{equation}
  \tilde X \wedge \tilde Y = (ad-bc) X\wedge Y\,.
\end{equation}
The quotient space of all parallelograms by this $SL(2,\reals)$ identification will be denoted by $A^2T_pM$,
and its elements are the oriented areas over $T_pM$. Similarly, we denote the bundle of oriented areas over $M$ by $A^2TM$.

Whereas parallelograms constitute a vector space $T_pM \oplus
T_pM$, the oriented areas $A^2T_pM$ at some point $p\in M$ cannot
carry a vector space structure. To see this note the following:
while $X \wedge Y$ clearly is an element of the vector space
$\Lambda^2T_pM$, a generic vector $\Omega \in \Lambda^2T_pM$
decomposes into a finite sum of such wedge-products, rather than
being a single wedge-product. A useful necessary and sufficient
criterion for $\Omega$ to be a simple wedge product, and hence an
oriented area, is
\begin{equation}\label{simple}
\Omega \wedge \Omega = 0\,,
\end{equation}
in which case $\Omega$ is called simple. Thus the space of oriented areas $A^2T_pM$ is a subset of the vector space $\Lambda^2T_pM$ defined by the vanishing of the four-tensor $\Omega\wedge\Omega$. Since this condition is quadratic in $\Omega$, the set $A^2T_pM$ is recognized as an affine variety in $\Lambda^2T_pM$. Hence the bundle $A^2TM$ fails to be a vector bundle. While this does not prevent the construction of a connection, from some connection on the underlying principal bundle, it is not possible to define a covariant derivative on $A^2TM$. But it is possible to define a covariant derivative on the vector bundle $\Lambda^2TM$, into which $A^2TM$ is embedded, and we will do so in section \ref{curvature}.

When discussing strings and string fluids in section \ref{sec_sfluids}, we will also need to address integrability issues, i.e., under which circumstances a distribution of oriented areas is tangent to some underlying two-surface.

\section{Abelian gauge fields and the Fresnel tensor}\label{gauge}
The physical postulate put forward in this paper is that physical
spacetime is an area metric manifold $(M,G)$,
rather than a metric manifold. This immediately prompts the
question of whether a generic area metric may give rise to some
effective metric on the manifold~$M$. The latter will play a significant role in the construction of an area metric curvature scalar which is downward compatible to the metric Ricci scalar, in section \ref{curvature}.

With the aim of constructing an effective metric in mind,
it turns out to be extraordinarily instructive to probe the
geometric structure of an area metric manifold by abelian gauge
theory. In particular this allows us to study wave propagation,
which in the geometric-optical limit gives insight into the
geometry of rays. In the metric-induced case, the ray surfaces reduce to the familiar null cones. In our more general setting, the Fresnel tensor describes the geometry of rays, and its derivation presents the first step towards the extraction of metric information from the underlying area metric manifold, which will be completed in sections \ref{normal} and \ref{sec_effmetric}.

Consider the following action for a one-form potential $A$ with field strength $F=dA$ on an area metric manifold of dimension $d\geq 3$:
\begin{equation}\label{edaction}
  -\frac{1}{2}\int_M \omega_G\,G^{-1}(F,F)=\int_M\omega_G\,L_G(F)\,.
\end{equation}
The equations of motion, derived by variation with respect to $A$, are simply given by the Bianchi identity $dF=0$ and by $dH=0$ in terms of the dual
electromagnetic induction~$H$, which is a $(d-2)$-form on $M$ with components
\begin{equation}\label{induction}
  H_{a_1 \dots a_{d-2}} = \omega_{G\,a_1 \dots a_{d-2} m n} \frac{\partial L_G(F)}{\partial F_{m n}}\,.
\end{equation}

To obtain geometric information from these equations, we
study the geometric-optical limit. In this limit one considers
waves as propagating discontinuities in the derivatives of the
fields $F$ and of the electromagnetic induction $(d-2)$-form $H$
along a wavefront surface described by the level lines of some
scalar function $\Psi$ on $M$. The rays of the wavefront are then
given by the gradient $p=d\Psi$. Extending an insightful argument
of Hehl, Obukhov and Rubilar \cite{Hehl:2004yk,Hehl:2005xu}, where a detailed
derivation may be found, to arbitrary dimension $d$, we find that
the wavefront gradients must obey the Fresnel equation
\begin{equation}\label{Fresneldens}
   \tilde{\mathcal{G}}^{a_1 \dots a_{2(d-2)}} p_{a_1} \dots p_{a_{2(d-2)}} = 0\,,
\end{equation}
where $\tilde{\mathcal{G}}$ is the totally symmetric tensor density of weight $-2$ given by
\begin{equation}\label{dense}
        \tilde{\mathcal{G}}^{a_1 \dots a_{2(d-2)}} = - \frac{(d-1)!}{4} \epsilon_{i i_1 \dots i_{d-1}} \epsilon_{j_1 \dots j_{d-1} j} C^{i i_1 j_1 (a_1} C^{a_2 | i_2 j_2 | a_3} \dots C^{a_{2(d-3)} | i_{d-2} j_{d-2} | a_{2d-5}} C^{a_{2(d-2)} ) i_{d-1} j_{d-1} j}\,,
\end{equation}
with $C$ being the cyclic part of the inverse area metric $G^{-1}$
as in (\ref{C+F}). As explained in appendix \ref{conventions} we use the convention that the summation over numbered \emph{anti}-symmetric indices is ordered, i.e., $i_1<i_2<\dots<i_{d-1}$ and similarly for the $j_k$.

It is a remarkable fact that, in the
geometric-optical limit, the propagation of wave fronts is
determined entirely in terms of the cyclic part of the inverse
area metric, due to (\ref{dense}). It will be convenient to cast the information encoded
in the density $\tilde{\mathcal{G}}$ into the form of a tensor
$\mathcal{G}$. In order to not introduce non-cyclic information
into this tensor, we are led to de-densitize $\tilde{\mathcal{G}}$
according to
\begin{equation}\label{Fresnel}
  \mathcal{G}^{a_1 \dots a_{2(d-2)}} = |\Det C|^{-1/(d-1)} \tilde{\mathcal{G}}^{a_1 \dots a_{2(d-2)}}\,.
\end{equation}
Note that the determinant factor presents the required scalar
density of weight $+2$ because~$C$ is a contravariant tensor. The
tensor $\mathcal{G}$ describing the ray surfaces will be called
the Fresnel tensor induced from the area metric $G$.

In order to illustrate the geometric role of the Fresnel tensor
$\mathcal{G}$ in a familiar setting, consider the action
(\ref{edaction}) on a Lorentzian manifold $(M,g)$, on which,
for the time being, the area
measure is simply induced from the metric $g$ by virtue of
(\ref{Cg}). A more subtle discussion of metric-induced measures
will follow in section \ref{sec_effmetric}. In this case the
action reduces to the standard form $\int F\wedge *_g F$.
Moreover, a direct calculation shows that
 the Fresnel tensor simplifies to
\begin{equation} \label{metricFresnel}
    \mathcal{G}^{a_1 \dots a_{2(d-2)}} = g^{(a_1 a_2} \dots g^{a_{2d-5} a_{2(d-2)})}\,,
\end{equation}
so that the Fresnel equation (\ref{Fresnel}) takes the factorized form
\begin{equation}\label{factoredFresnel}
     (g^{ab} p_a p_b)^{d-2} = 0\,.
\end{equation}
This result is of course equivalent to the standard null condition
for light rays propagating on a metric manifold.

The insight to be drawn from this comparison is the fact that, a priori, it is not a metric
that describes the propagation of rays (not even in the metric case),
but rather the Fresnel tensor $\mathcal{G}$ (which however happens to factorize neatly to the form
(\ref{factoredFresnel}) in the case of a metric background).
Hence, it is useful to make a conceptual difference between the background structure (metric or area metric)
employed to define one-form
dynamics on the one hand, and the structure that describes that theory in the geometric-optical limit (the Fresnel tensor)
on the other hand, although we are used to think of these structures as synonymous in the metric case.

There is an important corollary from the above findings, of which
we will make essential use in section \ref{sec_Killing}. For an
area metric in three dimensions, the (totally symmetric) Fresnel
tensor $\mathcal{G}^{ab}$ is of second rank, and with some amount
of algebra, one checks that $\mathcal{G}^{ab}$ is invertible.
Furthermore, $(\mathcal{G}^{-1})_{ab}$ induces the area metric we
started with, by expression (\ref{Cg}). Hence, in three dimensions,
every area metric is induced from some metric. This is no longer
the case in more than three dimensions.

\section{Normal area metrics and signature}\label{normal}
As a second step towards the extraction
of an effective metric from the area metric, we identify in
this section a family $h_\sigma$ of pre-metrics
on~$M$ which are induced by an
area metric via the Fresnel tensor. If the area metric data
distinguish a non-degenerate member of that family (in a way we
will explain below), the area metric manifold will be called
normal. In other words, normal area metrics allow for the
extraction of a metric.

Probing an area metric manifold of arbitrary dimension by abelian
gauge theory, we found that wave propagation in the
geometric-optical limit is described by the Fresnel tensor
(\ref{Fresnel}), not by a metric. We may use the totally symmetric
Fresnel tensor to define a symmetric bilinear form on a subbundle
of the cotangent bundle $T^*M$ (with bundle projection ${\pi:TM\rightarrow M}$) by the following construction, which in similar form finds application in Finslerian geometry \cite{Finslerbook}. We choose local
coordinates $x^a$ on $U \subset M$ which induce local coordinates
$(x^a,p_a)$ on $\pi^{-1}(U) \subset T^*M$, by virtue of $p = p_a
dx^a$. Then we define the characteristic function
\begin{equation}\label{character}
  h(x,p) := \left(\mathcal{G}(x)^{a_1 \dots a_{2(d-2)}} p_{a_1} \dots p_{a_{2(d-2)}}\right)^{1/(d-2)}
\end{equation}
on that portion $\mathcal{N}\subset T^*M$ of the cotangent bundle
where $h$ takes real values. Thus $\mathcal{N}$ is a level set of
the function on the total bundle space defined by the imaginary
part of $h$, and hence defines a subbundle of $T^*M$,
with $\pi(N)=M$. For any
$x\in M$, the function $h(x,p)$ is homogeneous of degree two in
$p$, i.e., $h(x,\lambda p) = \lambda^2 h(x,p)$.
From the characteristic function we may now define a
symmetric tensor field $h^{ab}$ by
differentiating~$h$ twice with respect to the fibre coordinate
\begin{equation}\label{tangentmetric}
h^{ab}(x,p) := \frac{1}{2} \frac{\partial^2}{\partial p_a\, \partial p_b} h(x,p)\,.
\end{equation}
This is indeed a tensor under diffeomorphisms of $M$ since they do not depend on the fibre coordinates $p$. It is quickly verified that
\begin{equation}\label{deg0}
h^{ab}(x,\lambda p) = h^{ab}(x, p)\,,
\end{equation}
so that the bilinear form $h$ only depends on the direction of~$p$ in the fibre, not on its `length'. This scaling behaviour of $h^{ab}$ is consistent with the scaling of the characteristic function $h$.

A simple and fruitful way of looking at this construction is to
think of an area metric as giving rise to a family $h_\sigma$ of
symmetric contravariant tensors on $M$ with components
\begin{equation}\label{hsigma1}
  h^{ab}_{\sigma(x)}(x) := h^{ab}(x,\sigma(x))\,,
\end{equation}
parametrized by sections $\sigma: M \to \mathcal{N}\subset T^*M$.
Viewing the characteristic function $h$ as a generalized norm squared on cotangent vectors $p$, the
symmetric covariant tensors $h_\sigma$ present linearizations
of this norm around the chosen section $\sigma$. We emphasize
again the fact that only the cyclic part $C$ of the inverse area
metric $G^{-1}$ contributed to the Fresnel tensor $\mathcal{G}$,
and hence to the construction of the symmetric
tensors $h^{ab}_\sigma$.

An important class of area metric manifolds are those for which
one may construct a particular section $\sigma_G: M \to
\mathcal{N}$ alone from the data of the area metric tensor $G$, such
that~$h^{ab}_{\sigma_G}$ is non-degenerate on all of $M$. Then
$(M,G,\sigma_G)$ will be called a normal area metric manifold.
We will see that normality of an area metric is an inevitable requirement for the construction of area metric curvature tensors that are downward compatible to their metric counterparts. But it is not a big restriction. We will show in the next section that one-forms $\sigma$ can always be constructed from the area metric, albeit not always uniquely. Also, $\mathrm{det}\,h^{ab}_{\sigma_G}\neq 0$ is an open condition so that normality is the rule not the exception. In particular, we will show that area metrics induced from Riemannian metrics are always normal. For Lorentzian metrics in even dimensions, we find that the construction of a normal area metric interestingly requires that the Lorentzian metric should describe a stably causal spacetime.

The signature of a normal area metric manifold $(M,G,\sigma_G)$ is
defined as the signature $(t,s)\equiv(-\dots - +\dots +)$ of the
metric $h_{\sigma_G}$ constructed from $G$ and $\sigma_G$.
With this definition, we may define the totally antisymmetric
tensor density $\epsilon^{a_1 \dots a_d}$ to be normalized such
that $\epsilon^{1\dots d}=(-1)^t$. This in turn induces the
contravariant tensor $\overline\omega$ with components
\begin{equation}
    \overline\omega_G^{a_1 \dots a_d} = |\mathrm{Det}\,G|^{-1/(2d-2)} \epsilon^{a_1 \dots a_d}\,.
\end{equation}
It is necessary to distinguish the tensors $\omega$ and
$\overline\omega$, because their components are in general not
related by lowering and raising of indices with the area metric
and inverse area metric, respectively. With these definitions, we
obtain for the contraction over $k$ indices the useful identity
\begin{equation}
   \overline\omega_G^{a_1\dots a_{d-k}\, m_1 \dots m_k}\, \omega_{G\,\,}{}_{b_1 \dots b_{d-k} \, m_1 \dots m_k} = (-1)^t\, (d-k)!\, \delta^{[a_1 \dots a_{d-k}]}_{\,\,\,b_1 \dots b_{d-k}}\,.
\end{equation}

\section{Effective metric}\label{sec_effmetric}
We now complete the construction of an effective metric
from a generic area metric manifold $(M,G)$ without specifying
additional data. From our findings in sections~\ref{gauge}
and~\ref{normal}, we know that the extraction of a metric from
area metric data amounts to the construction of a section $\sigma$
of the subbundle $\mathcal{N}\subset T^*M$, as defined by reality
of the characteristic function~(\ref{character}), such that~$(M,G,\sigma)$
is normal. Then the inverse of the
tensor~$h_\sigma$ as defined
in~(\ref{hsigma1}) provides the desired metric. In order to achieve
downward compatibility with metric geometry, we further require
that for metric-induced area metrics the inducing metric~$g$ is
recovered up to a sign (more cannot be expected since the metric-induced area metrics are quadratic expressions).
This requirement will lead to a modification of the
induction formula~(\ref{Cg}) for Lorentzian metrics.

The Fresnel tensor for a metric-induced area metric (\ref{Cg}) takes the simple form (\ref{metricFresnel}). Then the characteristic function (\ref{character}) is given by
\begin{equation}
   h(x,p) = (g^{-1}(p,p)^{d-2})^{1/(d-2)} = \left\{\begin{array}{ll} |g^{-1}(p,p)| & \textrm{if } \dim M \textrm{ even}\\ g^{-1}(p,p) & \textrm{if } \dim M \textrm{ odd}\end{array}\right..
\end{equation}
In both cases this is real on the entire cotangent bundle, so that $\mathcal{N}=T^*M$. Now, however, appears a crucial difference between metric manifolds of different signatures. First consider the case of a Riemannian metric $g$ where $h(x,p)=g^{-1}(p,p)$ independent of the dimension. In this case all members of the family $h^{ab}_\sigma$ of metrics coincide, and are given, for an arbitrary section $\sigma$, by
\begin{equation}
  h_\sigma^{ab}(x) = g^{ab}(x)\,.
\end{equation}
This means that any choice of section $\sigma(x)$ recovers the inducing Riemannian metric $g$ from the area metric $G_g$, including the zero section that does not even require area metric data. Thus $(M,G_g)$ is normal if $g$ is a Riemannian metric. The same conclusion holds if $g$ is an odd-dimensional Lorentzian metric. But now let $g$ be an even-dimensional Lorentzian metric. Then the family $h_\sigma$ of metrics is given by
\begin{equation}\label{Lorentzh}
  h_\sigma^{ab}(x) = \frac{g^{-1}(\sigma,\sigma)}{|g^{-1}(\sigma,\sigma)|} g^{ab}(x)\,.
\end{equation}
In contrast to the Riemannian case, it is now not possible to
choose the zero-section in order to recover the inducing
Lorentzian metric.

This means we must obtain a non-trivial section from the area
metric in order to be downwards compatible to the metric-induced
case. Otherwise the construction would fail for the
even-dimensional Lorentzian case. We first observe that the
required section $\sigma$ cannot be constructed alone from the
cyclic part $C$ of the inverse area metric. This is because the
only one-forms obtainable from $C$ are of the form
\begin{equation}\label{candidateone}
  d\, \textrm{Tr} \prod_i (C*)^{n_i} (*C^{-1})^{m_i}
\end{equation}
where $n_i$ and $m_i$ are a finite set of non-negative integers
and $(C*)^{ab}{}_{cd} = C^{abm_1m_2}\omega_{C\, m_1m_2cd}$ and
$(*C^{-1})^{ab}{}_{cd} = \overline\omega_C^{abm_1m_2}
C_{m_1m_2cd}$. Now for an area metric that is metric-induced
according to~(\ref{Cg}), we have $C=C_g^{-1}$ and $C^{-1}=C_g$, so
that both $C$ and its inverse are cylic (which, we recall, does
not hold for generic area metrics). Then it is easily verified
that all of the one-forms~(\ref{candidateone}) vanish. Thus we are
compelled to conclude that information from the area metric beyond
the cyclic part~$C$ is needed for the definition of a non-trivial
section~$\sigma$. We hence turn to the four-form part of the
inverse area metric as a potential carrier of this information.
This fits nicely together with the observation that~$h^{ab}(x,p)$
is defined entirely in terms of the cyclic part of the inverse
area metric. So the four-form part will exclusively contribute to
the construction of the section~$\sigma$ which replaces~$p$, while
the cyclic part exclusively contributes to the definition of the
bilinear symmetric tensor~$h^{ab}(x,p)$.

For an area  metric manifold of dimension $d\geq 4$, we may
schematically decompose the inverse area metric as
\begin{equation}\label{scheme}
  G^{-1} = C \, + \, \overline\omega_C\, \llcorner\, \phi
\end{equation}
so that the information in the four-form part is succinctly
encoded in a $(d-4)$-form $\phi_{(d-4)}$. Now four-dimensional
area metric manifolds are distinguished since, in this dimension,
the construction of a one-form from the four-form part of $G^{-1}$
is unique. The only possible choice is
\begin{equation}
  \sigma_G = d\phi_{(0)}.
\end{equation}
Any rescaling of this choice, even by a function, is without
effect because of relation (\ref{deg0}). Due to the symmetries of
$G^{-1}$ and $\overline\omega$, it is also not possible to
construct further non-vanishing one-forms from $d\phi$. Already in
five dimensions, there are several possibilities,
\begin{equation}
  \phi_{(1)}\,,\qquad d C(d\phi_{(1)},d\phi_{(1)})\,.
\end{equation}
In even higher dimensions, the number of possibilities further proliferates.

So in four dimensions only may one speak of \emph{the} normal area
metric manifold $(M,G,\sigma)$ constructed from an area metric
manifold. (At least, unless one could
formulate further meaningful criteria to construct, also in higher dimensions, a natural section $\sigma_G$  to single out the effective metric from the
family $h_\sigma$ of contravariant tensors.) From now on, we will therefore
restrict attention to the four-dimensional case, which of course
also is the one of immediate physical interest. A four-dimensional
normal area metric manifold will henceforth be denoted $(M,G,g_G)$ where
\begin{equation}\label{zzz}
  g_G = h_{\sigma_G}= h_{d\phi_{(0)}}
\end{equation}
is the unique metric constructed from the area metric~$G$.
Generically, the metric $g_G$ of course only encodes part of
the information contained in the area metric. Other than in
dimension three, the area measure $G_{g_G}$ re-induced from the
effective metric generically does not agree with the measure
determined by~$G$.

The above constructions play out nicely for area metric manifolds
induced by stably causal spacetimes $(M,g)$ in four dimensions.
For the latter, there always exists a global time function $\phi$
on $M$ such that $d\phi$ is everywhere $g$-timelike \cite{Beem}.
Hence after a choice of global time function $\phi$ has been made,
the latter may be encoded directly into the totally antisymmetric
part of the inverse area metric:
\begin{equation}\label{stablycausallift}
   (G_g^\phi)^{-1} = C_g^{-1} + \phi\, \overline\omega_{g}\,.
\end{equation}
Due to our mainly plus signature convention, the inducing
Lorentzian metric may be recovered, using (\ref{zzz}) and
(\ref{Lorentzh}), as $g=-g_G$. So downward compatibility is
maintained. From now on, we will denote the inverse area metric
(\ref{stablycausallift}) induced by some Lorentzian metric $g$ and
a suitable scalar field $\phi$ always by $(G_g^\phi)^{-1}$.
Inverting this one finds the relation
\begin{equation}\label{invlift}
G_g^\phi=\left(1+\phi^2\right)^{-1}\left(C_g-\phi\omega_g\right).
\end{equation}
This is the closest a normal area metric may come to a Lorentzian
metric, and presents an important class of area spacetimes which
can be easily compared to standard metric spacetime, see section
\ref{solar}. In the following section, we will see that
cosmological symmetries, for instance, enforce this special,
almost metric, form of an area metric.

\section{Isometries and Killing vectors}\label{sec_Killing}
At this point it is useful to study isometries of area metric
manifolds. This will provide us with more evidence for the natural
role non-cyclic area metrics play and
prepare our discussion of area metric cosmology.

A diffeomorphism $h: M \to M$ is called an area metric isometry if
it preserves the area metric in the sense that for all smooth
vector fields $U,V,A,B$  on $M$ and all $p\in M$, we have
\begin{equation}
  G_{h(p)}(h_*U, h_*V, h_*A, h_*B) = G_p(U,V,A,B)\,,
\end{equation}
where $h_*$ denotes the push-forward with respect to $h$. As in
the metric case, it is useful to consider the generators of
isometries, i.e., Killing vector fields $X$. Clearly, $X$ is a
Killing vector field for an area metric manifold $(M,G)$ if
$\mathcal{L}_X G = 0$. This condition implies that the Killing
vectors of a given area metric manifold, together with the
standard commutator, constitute a Lie algebra.

It is reassuring to verify that for a metric-induced area metric
manifold, $X$ is a Killing vector of $(M,C_g)$ if and only if $X$
is a Killing vector of $(M,g)$. This is indeed the case: the
implication ${\cal L}_X g=0\Rightarrow {\cal L}_X C_g=0$ is
evident. Conversely, assume ${\cal L}_X C_g=0$. In particular, for
any pair of $g$-orthogonal vectors $A, B$, we have
\begin{equation}\label{Killing}
0=({\cal L}_X C_g)(A,B,A,B)=({\cal L}_X g)(A,A)g(B,B)+({\cal L}_X
g)(B,B)g(A,A)\,.
\end{equation}
Now consider a $g$-orthonormal basis $\{A_i\}$, $i=1,\dots,d$ of
$T_p M$. For any three distinct vectors $A, B, C$ in such a
basis, the relation above implies
\begin{subequations}
\begin{eqnarray}
 0&=&({\cal L}_X g)(A,A)g(B,B)+({\cal L}_X g)(B,B)g(A,A)\,,\\
  0&=&({\cal L}_X g)(A,A)g(C,C)+({\cal L}_X g)(C,C)g(A,A)\,,\\
   0&=&({\cal L}_X g)(B,B)g(C,C)+({\cal L}_X g)(C,C)g(B,B)\,.
\end{eqnarray}
\end{subequations}
This set of equations is homogeneous and non-degenerate with
respect to $({\cal L}_X g)(A,A)$, $({\cal L}_X g)(B,B)$, and $({\cal
L}_X g)(C,C)$, so that it admits as the unique solution $({\cal
L}_X g)(A,A)=({\cal L}_X g)(B,B)=({\cal L}_X g)(C,C)=0$. To
complete the proof, note that we then also have
\begin{equation}
0=({\cal L}_X C_g)(A,C,B,C)=({\cal L}_X g)(A,B)g(C,C)\,,
\end{equation}
so that $({\cal L}_X g)(A,B)=0$. It follows that ${\cal L}_X g
(A_i,A_j)=0$ for all $i,j=1, \dots, d$; in other words, ${\cal
L}_X g=0$. It is also worthwhile to note that, in the case
$G=C_g+\omega\llcorner\phi$, the theorem easily extends to ${\cal
L}_X G=0\Leftrightarrow {\cal L}_X g=0 \wedge {\cal L}_X\phi=0$.

Equipped with the definition of Killing vectors, we can now
discuss physically relevant applications. In particular, we
will be able to construct the analogue of a
Friedmann-Lema\^itre-Robertson-Walker (FLRW) cosmology for an area
metric background. We will discover that a cosmological area
metric in $d=4$ is precisely of the form (\ref{stablycausallift}),
with the cyclic part being induced by a standard FLRW metric, and
with the scalar field providing an additional degree of freedom in
comparison to metric geometry.

First recall that, given a Lie group action $G \times M \to M$ on
some smooth manifold $M$, it is useful to define the following
notions \cite{Isham}. The orbit of some point $p\in M$ is the
submanifold ${\cal O}_p=\{q\in M\,|\,q=g.p\,\, \textrm{for some}\,\,
g\in G\}$. The isotropy group $I_p$ at some point $p\in M$
is that subgroup of $G$ whose action leaves $p$ invariant. A
submanifold $N$ of dimension $n$ is said to be \emph{homogeneous}
if there is a transitive action of some Lie group $G$ on $N$, i.e.
if ${\cal O}_p= N$ for all $p\in N$. It is said to be
\emph{spherically symmetric} around a point $p$ if the isotropy
group of $p$ is $SO(n)$, and the relative orbit of any other point $q\neq p$ around $p$ is topologically equivalent to an $(n-1)$-sphere.

In complete analogy with the standard, rigorous definition of a
(spatially) homogeneous and isotropic manifold in Lorentzian
geometry, we can now define area metric cosmology, i.e., FLRW area metric
manifolds, as follows: $(M,G)$ is a $d$-dimensional FLRW
area metric manifold if and only if it can be suitably sliced in
$(d-1)$-dimensional hypersurfaces, which are homogeneous
 and spherically symmetric around
each point. Additionally, the
hypersurfaces must be spacelike, in the sense that the restriction
of the area metric to them must be positive definite.

In the four-dimensional case, on which we will focus, the general
form of the area metric compatible with these requirements can be
easily determined. First of all, we note that the area metric on
the slices must be metric-induced, since the slices are
three-dimensional (compare the corollary at the end of section
\ref{gauge}). The inducing metric $\bar g$ must describe three-dimensional
maximally symmetric Riemannian manifolds with line element
\begin{equation}\label{FRW}
\bar g_{\alpha\beta}dx^{\alpha}dx^{\beta}=
\frac{dr^2}{1-kr^2}+r^2\left(d\theta^2+\sin^2\theta d\phi^2\right)
\end{equation}
for normalized curvature $k=0,\pm 1$, in the usual
system of coordinates $\{x^0,x^{\alpha}\}=\{t,r,\theta,\phi\}$.
For the restriction of the area metric to the slices
of constant $t$ we then obtain the expression
$G_{|t\,\alpha\beta\gamma\delta}=S^2(t)C_{\bar{g}\,\alpha\beta\gamma\delta}$
for some function $S(t)$. Moreover, on each of the spacelike slices,
$G_{0\alpha 0\beta}$ behaves as a three-dimensional metric with
respect to the entire group of isometries. Therefore, introducing some
function $Q(t)$, we have
\begin{equation}
 G_{0\alpha 0\beta}=Q(t)\bar{g}_{\alpha\beta}\,.
\end{equation}
Finally, explicitly solving Killing conditions for the remaining
components of the area metric, one finds for the remaining
non-vanishing components the form
\begin{equation}
G_{tr\theta\phi}=-G_{t\theta r\phi}=G_{t\phi
r\theta}=F(t)\frac{r^2\sin\theta}{\sqrt{1-kr^2}}
\end{equation}
for some function $F(t)$. Employing a suitable rescaling of the
time coordinate and redefining the free functions
$S(t),Q(t),F(t)$, it is easy to verify that an FLRW area metric
takes the form $G=G_{\hat g}^\Phi$, see (\ref{stablycausallift}),
where $\hat g$ is the standard FLRW metric, so that
\begin{equation}
\hat{g}_{ab}dx^a dx^b=-dt^2+a^2(t)\bar
{g}_{\alpha\beta}dx^{\alpha}dx^{\beta}\,,
\end{equation}
and $\Phi=\Phi(t)$ only depends on time. Inverting $G=G_{\hat g}^\Phi$ and redefining the scalar field, also the inverse area metric decomposes into the form
\begin{equation}
  (G_{\hat g}^\phi)^{-1}=C^{-1}_{\hat{g}}+\phi\overline{\omega}_{\hat g}\,.
\end{equation}

Thus in comparison with the standard metric case, FLRW area metric
manifolds in $d=4$ feature a time-dependent scalar degree of freedom $\phi(t)$
additional to those of the metric, which happens to be encoded in
the totally antisymmetric part of the (inverse) area metric. If
non-vanishing, the section $\sigma=d\phi$ will be $\hat
g$-timelike, and thus render the FLRW area metric normal,
as discussed at the end of the previous section.
Consequences of this feature for the dynamics of such cosmological
models will be discussed in sections~\ref{sec_cosmology} and~\ref{sec_exact}.

\section{Area metric connection and curvature}\label{curvature}
Our aim in this section is the construction of an area metric
compatible connection and area metric curvature, which are downward compatible to their metric counterparts.
As discussed in \cite{Schuller:2005ru}, we
principally need a connection on the non-vector bundle of area
spaces $A^2TM$ embedded in~$\Lambda^2TM$. In order to keep
technicalities to a minimum, we will determine this connection in
terms of a covariant derivative $\nabla$ on the vector bundle
$\Lambda^2TM$.

Consider a four-dimensional normal area metric manifold $(M,G)\sim
(M,G,g_G)$, where~$g_G$ is the unique metric constructed from $G$,
as explained in section \ref{sec_effmetric}. The metric $g_G$
immediately gives rise to the torsion-free Levi-Civita connection
$\nabla^{LC}$, which of course lifts to the $\Lambda^2TM$-bundle
in the standard way. While $\nabla^{LC}$ provides a covariant
derivative on $\Lambda^2TM$ and is downward compatible, it is not of direct relevance for
the area metric manifold  $(M,G)$. On the one hand, it does not
contain all the information contained in the area metric, which
follows from a simple counting of the number of independent
components of the effective metric that defines $\nabla^{LC}$. On
the other hand, $\nabla^{LC}$ does not obey the important
geometric property that the area metric should be covariantly
constant: $\nabla^{LC} G\neq 0$ in general. But covariant
constancy of the area metric ensures, for instance, that the simplicity
condition~(\ref{simple}) of a section~$\Omega$ of~$\Lambda^2TM$
is not violated by parallel transport; in other words,
areas are preserved under parallel transport~\cite{Schuller:2005ru}.
But most importantly, we require area metric compatibility because
it allows us to determine a unique connection as follows.

First note that an arbitrary connection $\nabla$ on $\Lambda^2TM$,
including the area metric compatible one we are seeking, differs
at most by some tensor $X$ from the lift of the Levi-Civita
connection, i.e.,
\begin{equation}\label{Xdef}
  \nabla_Z\Omega=\nabla_Z^{LC}\Omega+X(Z,\Omega)
\end{equation}
or, in coordinates, $(\nabla_f \Omega)^{ab} = (\nabla^{LC}_f \Omega)^{ab} + X^{ab}{}_{i_1i_2 f} \Omega^{i_1i_2}$. In order to uniquely determine an area metric-compatible
connection $\nabla$, we need to constrain the class of
$\Lambda^2TM$-connections we are looking for. To this end note
that for an arbitrary connection on the bundle $\Lambda^2TM$ over
a normal area metric manifold $(M,G,g_G)$, we may define the
tensor
\begin{equation}
  T_G(Z,\Omega,\Sigma) = G((\nabla_Z-\nabla_Z^{LC})\Omega,\Sigma) - G(\Omega,(\nabla_Z-\nabla_Z^{LC})\Sigma)\,,
\end{equation}
for any vector $Z$ and sections $\Omega, \Sigma$ of $\Lambda^2TM$.
We call $T_G$ the relative torsion of the
$\Lambda^2TM$-connection~$\nabla$ with respect to $(M,G,g_G)$. The
virtue of having such a tensor is that one may formulate a
consistent condition on a generic $\Lambda^2TM$ connection by
requiring $T_G$ to vanish identically. This requirement corresponds
to the symmetry condition $X_{abcdf} = X_{cdabf}$ on the components
of the tensor $X$ defined in (\ref{Xdef}), where indices have been
lowered using the area metric $G$.
Vanishing relative torsion and area metric compatibility together,
\begin{equation}
  \nabla G = 0 \qquad \textrm{ and } \qquad T_G = 0\,,
\end{equation}
then uniquely determine the tensor $X$: using the first requirement $\nabla G=0$ gives
\begin{equation}
-(\nabla_Z^{LC}G)(\Omega,\Sigma)=(\nabla_Z G-\nabla_Z^{LC}G)(\Omega,\Sigma)=-G(X(Z,\Omega),\Sigma)-G(\Omega,X(Z,\Sigma))\,,
\end{equation}
while the second requirement $T_G=0$ allows us to equate this to $-2G(X(Z,\Omega),\Sigma)$. Then $X$ must be of the form
\begin{equation}\label{XX}
X(Z,\Omega)=\frac{1}{2}G^{-1}((\nabla_Z^{LC}G)(\Omega,\cdot),\cdot)\,,\qquad X^{ab}{}_{cd f} = \frac{1}{4}G^{abij} \nabla^{LC}_f G_{ijcd}\,.
\end{equation}

This result can also be obtained in the coordinate-free
language of pre-connections introduced in \cite{Schuller:2005ru}. Then the
relative torsion-free, area metric compatible
$\Lambda^2TM$-connection is given by
\begin{equation}
  \nabla_Z \Omega = \frac{1}{2} G^{-1}(D_Z(\Omega,\cdot) + D_Z[\Omega,\cdot],\cdot)\,,
\end{equation}
with the symmetric and antisymmetric pre-connections
\begin{subequations}
\begin{eqnarray}
  D_Z(\Omega,\Sigma) &=& Z G(\Omega,\Sigma)\,,\\
  D_Z[\Omega,\Sigma] &=& G(\nabla_Z^{LC}\Omega,\Sigma) - G(\Omega,\nabla_Z^{LC}\Sigma)\,,
\end{eqnarray}
\end{subequations}
respectively. Here the symmetric pre-connection is determined by $\nabla G=0$ alone, while the antisymmetric pre-connection then is determined by $T_G=0$. Note that while in~\cite{Schuller:2005ru} a metric~$g$ was
assumed to be given as data in addition to an area metric in order
to construct a connection, the significant advance of this paper
consists in having constructed the metric~$g_G$ in a unique manner
from the area metric data alone. Hence we have now achieved the
construction of a true area metric connection, without additional
data.

The identification of this unique connection $\nabla$ associated
with a normal four-dimensional area metric manifold $(M,G)$ now
allows for the definition of the area metric curvature in standard
fashion
\begin{equation}\label{areacurv}
\mathcal{R}_G(X,Y)\Omega=\nabla_X\nabla_Y\Omega-\nabla_Y\nabla_X\Omega-\nabla_{[X,Y]}\Omega\,,
\end{equation}
whose components can be calculated to be
\begin{equation}
  \mathcal{R}_G{}^{a_1a_2}{}_{b_1b_2ij}=4\delta^{[a_1}_{[b_1}R^{a_2]}{}_{b_2]ij}+\left(\nabla^{LC}_i X^{a_1a_2}{}_{b_1b_2j}+X^{a_1a_2}{}_{f_1f_2i}X^{f_1f_2}{}_{b_1b_2j}-(i\leftrightarrow j)\right),
\end{equation}
where $R$ is the curvature of the Levi-Civita connection
$\nabla^{LC}$ obtained from $g_G$.
Natural contraction over indices 1 and 2 with
indices 3 and 5 yields the area metric Ricci tensor
\begin{equation}
  \mathcal{R}_G{\,}_{mn} = \mathcal{R}_G{}^{pq}{}_{pmqn}\,.
\end{equation}
Another second rank tensor is defined by
\begin{equation}
  N_{[ij]} = \mathcal{R}_G{}^{pq}{}_{pqij}\,.
\end{equation}
Due to its antisymmetry, however, there is no contraction to a
non-zero scalar of first order in~$N$; the lowest order scalar one
may build from $N_{ij}$ is $G^{-1}(N,N)$. This invariant is
therefore of no interest to area metric gravity, as its inclusion
in an action would drive the derivative order of the equations of
motion automatically beyond two. So the unique area metric
curvature scalar of linear order in
$\mathcal{R}_G^{a_1a_2}{}_{b_1b_2ij}$, is the area metric Ricci
scalar
\begin{equation}
  \mathcal{R}_G = g_G^{mn}\,\mathcal{R}_G{}_{mn}\,.
\end{equation}

In the case of an area metric
$C_g$ induced from a Riemannian metric we have~$X=0$ from~(\ref{XX}).
Hence it is immediately clear that the area metric Ricci
tensor and Ricci scalar reduce to their metric counterparts while
the tensor $N$ vanishes. The Lorentzian case is less obvious;
recall that inducing a normal area metric from a Lorentzian metric
requires the addition of a scalar $\phi$ whose gradient $d\phi$ is
a globally non-null one-form field; according to~(\ref{stablycausallift})
the area metric is~$G^{-1}=(G_g^{\phi})^{-1}$.
It may now be checked, however, that
also in this case all of the above curvature tensors reduce to
their metric counterparts, and that~$N$ vanishes identically. This
proves the required downward compatibility to metric geometry. We will use
the curvature invariant~$\mathcal{R}_G$
in the following second part of the paper for the construction
of an area metric gravity theory.

\newpage
\section*{P A R T  \quad T W O :\\A R E A \quad  M E T R I C \quad G R A V I T Y}
The discussion of area metric gravity in this second part is self-contained, since the relevant results of area metric geometry developed in Part One are concisely summarized in the following section \ref{synopsis}. The general formalism of area metric gravity and its coupling to matter is developed in section \ref{gravtheory}, followed by a comparison of almost metric vacuum solutions to standard general relativity in section \ref{solar}. Fluids on area metric spacetimes are necessarily string fluids, as is explained in sections \ref{pfluids} and \ref{sec_sfluids}. Building on these preliminaries, we apply area metric geometry to cosmology in section \ref{sec_cosmology}, and find the exact solution for a homogeneous and isotropic area metric universe filled with string dust in section \ref{sec_exact}. This leads to the prediction that area metric cosmology may explain, without dark energy or fine-tuning, the small late-time acceleration of our Universe.

\section{Practical synopsis of area metric geometry}\label{synopsis}
In this brief section, we give a practical guide to the construction of area metric curvature in \emph{four} dimensions. For derivations and details of the construction in coordinate-free fashion, see Part One of this paper. A Lorentzian area metric on a smooth four-dimensional manifold~$M$ is a covariant tensor
\begin{equation}
  G_{abcd} = G_{[ab][cd]} = G_{[cd][ab]}\,,
\end{equation}
whose inverse is defined as the contravariant tensor $G^{abcd}$ satisfying
\begin{equation}
  G^{abmn} G_{mncd} = 4 \delta^{[a}_c\delta^{b]}_d\,.
\end{equation}
Considering ordered pairs of indices $a_1 < a_2$, one may write an area metric in four dimensions in Petrov notation, i.e., as a $6\times 6$ matrix $G_{AB}$ with $A, B$ taking values in the set $\{[01], [02], [03], [12], [13], [23]\}$. Defining $\textrm{Det } G$ as the determinant taken over the matrix~$G_{AB}$, one may define covariant and contravariant area metric volume tensors
\begin{equation}
  \omega_{G\,abcd} = |\textrm{Det } G|^{1/6} \epsilon_{abcd} \quad \textrm{and} \quad \omega_G^{abcd} = |\textrm{Det } G|^{-1/6} \epsilon^{abcd}\,,
\end{equation}
where the totally antisymmetric tensor densities $\epsilon$ are normalized so that $\epsilon_{0123}=+1$ and ${\epsilon^{0123}= -1}$, respectively.
The inverse area metric uniquely decomposes as
\begin{equation}
   G^{abcd} = C^{abcd} + \phi \omega_C^{abcd}\,,
\end{equation}
i.e., into a cyclic part $C^{a[bcd]}=0$ and a totally antisymmetric part, which in four dimensions amounts to the specification of a function~$\phi$ on~$M$. The definition of~$\omega_C$ is that of~$\omega_G$, with~$C$ replaced by $G$. A normal area metric allows to extract a unique effective metric
\begin{equation}
  g_G^{ab}(x) = \frac{1}{2} \left.\frac{\partial^2}{\partial p_a \partial p_b}\right|_{p=d\phi}\left(-\frac{1}{24}\omega_{C\,mnpq}\omega_{C\,rstu}C^{mnri}C^{jpsk}C^{lqtu} p_i p_j p_k p_l \right)^{1/2}\,,
\end{equation}
from the area metric data. In case the area metric is almost metric, i.e.,
\begin{equation}
G^{abcd}=g^{ac}g^{bd}-g^{ad}g^{bc} + \phi\omega_g^{abcd}
\end{equation}
for some metric $g$, the effective metric $g_G$ recovers the inducing metric $g$ up to a sign,
which validates the present construction independent of its deeper geometric meaning discussed in Part One of this paper. Finally, the area metric curvature tensor is given by
\begin{equation}
  \mathcal{R}_G^{[a_1a_2]}{}_{[b_1b_2][ij]} = 4\delta^{[a_1}_{[b_1}R^{a_2]}{}_{b_2]ij}+\Big(\nabla^{LC}_i X^{a_1a_2}{}_{b_1b_2j}+\frac{1}{2}X^{a_1a_2}{}_{pqi}X^{pq}{}_{b_1b_2j}-(i\leftrightarrow j)\Big),
\end{equation}
where $R$ and $\nabla^{LC}$ are the Riemann tensor and the Levi-Civita connection of the effective metric $g_G$, and the non-metricity tensor $X$ is defined by
\begin{equation}
  X^{a_1a_2}{}_{b_1b_2 f} = \frac{1}{4} G^{a_1a_2mn} \nabla_f G_{mnb_1b_2} = X^{[a_1a_2]}{}_{[b_1b_2]f}\,.
\end{equation}
The area metric curvature tensor, as well as the associated area metric Ricci tensor ${(\mathcal{R}_G)_{ab} = \mathcal{R}^{pq}{}_{paqb}}$ and area metric Ricci scalar $\mathcal{R} = g_G^{ab}(\mathcal{R}_G)_{ab}$, reduce to their metric counterparts for almost metric area metrics. These correspondences ensure in particular that area metric geometry is downward compatible to metric geometry, which is therefore contained as a special case.

\section{Area metric gravity action and matter coupling}\label{gravtheory}
Area metric geometry allows us to devise a gravity theory different
from Einstein's without modifying the form of the Einstein-Hilbert
action; the latter is just re-interpreted as an action for an area
metric manifold, with all metric quantities being replaced by
their area metric counterparts. We restrict our study of the area
metric version of the Einstein-Hilbert action to four dimensions,
as only there the standard metric version enjoys its truly special
status due to Lovelock's theorem \cite{Lovelock:1971yv,Lovelock:1972vz}.
Thus we adopt the following action for four-dimensional normal area
metric spacetimes $(M,G)$:
\begin{equation}\label{totalaction}
 S_{grav}\, + \,S_m \, = \frac{1}{2\kappa}\int_M\omega_G\,\mathcal{R}_G \, + \, \int_M \mathcal{L}_m
\end{equation}
from which the equations of motion are derived by variation with
respect to the (inverse) area metric $G$. The gravitational constant $\kappa$ is not yet determined at this stage. The quantity $\mathcal{L}_m$
represents the  matter Lagrangian scalar density, which in many cases is simply $\mathcal{L}_m=\omega_GL_m$ in terms of a scalar Lagrangian $L_m$.

The diffeomorphism invariance of the action immediately implies an
area metric Bianchi identity for the gravitational part of the
action, and the conservation law for the area metric
energy-momentum tensor $T$. In order to study the latter, consider
a diffeomorphism on~$M$ generated by a vector field $\xi$. The
induced variation of the inverse area metric may be expressed
covariantly as a Lie derivative,
\begin{equation}
\delta G^{abcd}=(\mathcal{L}_\xi G)^{abcd}=\xi^p\nabla^{LC}_p G^{abcd}+2 \nabla^{LC}_p \xi^{[a}G^{b]pcd}+2 \nabla^{LC}_p \xi^{[c}G^{d]pab}\,,
\end{equation}
where the second equality uses the Levi-Civita connection of the
effective metric. Requiring invariance of the matter action under
this variation by setting $\delta S_m=0$, i.e., $0=\int_M\delta
G^{abcd}\,\delta S_m/\delta G^{abcd}$, and performing
some partial integration, we are led to the conservation equation
\begin{equation}\label{conservation}
T_{abcd}\nabla^{LC}_i G^{abcd}-4\left(\nabla^{LC}_p+\frac{1}{2(d-1)}X_p\right)\left(T_{abcd}G^{abp[c}\delta^{d]}_i\right)=0\,.
\end{equation}
for the fourth rank tensor
\begin{equation}
T_{abcd}=-|\mathrm{Det}\,G|^{-1/(2d-2)}\frac{\delta S_m}{\delta
G^{abcd}}\,.
\end{equation}
We call this tensor $T$ the generalized energy-momentum tensor of
matter on an area metric manifold. Since it is derived by
variation with respect to the inverse area
metric, it has the symmetries of the area metric. In particular it
may contain a totally antisymmetric part in the decomposition
under the local frame group.

Computationally performing the variation of the gravitational part
of the action is rather involved; it is given in full detail in
appendix \ref{appvary}. From the reducibility of the inverse area
metric $G$ into a cyclic and totally antisymmetric part, however,
it is clear that in four dimensions the equations of motion for
generic area metrics can be separated accordingly. So if
the tensor $K_{abcd}$ denotes the variation of the gravitational part of the
action~(\ref{totalaction}) with respect to the inverse area metric
$G^{-1}$, i.e.,
\begin{equation}
K_{abcd}=|\mathrm{Det}\,G|^{-1/6}\frac{\delta(\kappa S_{grav})}{\delta
G^{abcd}}\,,
\end{equation}
then, using equation (\ref{varG}), we may decompose the total variation into those terms arising from the gravitational part
\begin{equation}
\delta S_{grav}=\frac{1}{\kappa}\int_M\omega_G\left[\delta\phi\left(\bar\omega_C^{abcd}K_{abcd}\right)+\delta C^{abcd}\left(K_{abcd}+\frac{1}{24}\phi\bar\omega_C^{ijkl}K_{ijkl}C^{-1}_{abcd}\right)\right],
\end{equation}
and into those from the matter part
\begin{equation}
\delta S_m=-\int_M\omega_G\left[\delta\phi\left(\bar\omega_C^{abcd}T_{abcd}\right)+\delta C^{abcd}\left(T_{abcd}+\frac{1}{24}\phi\bar\omega_C^{ijkl}T_{ijkl}C^{-1}_{abcd}\right)\right].
\end{equation}
Taking care to impose the symmetries of
$\delta C^{abcd}$ on the expressions in brackets that are contracted with
it, we may thus define the cyclic contributions $K^C$ and $T^C$,
and the scalar contributions $K^\phi$ and $T^\phi$ of the
gravitational tensor $K$ and the energy-momentum tensor $T$ to the
equations of motion. In terms of these quantities, whose explicit
form is derived in appendix \ref{appvary}, the full equations of
motion for area metric gravity take the form
\begin{subequations}\label{generalform}
\begin{eqnarray}
   K^C_{abcd} &=& \kappa T^C_{abcd}\,,\label{generalform1}\\
   K^\phi &=& \kappa T^\phi\,.\label{generalform2}
\end{eqnarray}
\end{subequations}
Of course, these field equations are far too complex to be
exactly solved in general, and one has to content oneself with
studying exact solutions for highly symmetric spacetimes. For
area metric cosmology we will start this program below.

As an example for the general theory outlined above, consider the theory of electrodynamics defined by the action (\ref{edaction}). Variation of this action with respect to the inverse area metric yields the generalized energy momentum tensor
\begin{equation}\label{edT}
T_{abcd}=\frac{1}{8}F_{ab}F_{cd}-\frac{1}{64(d-1)}G_{abcd}G^{ijkl}F_{ij}F_{kl}\,.
\end{equation}
As an aside we note here that $d=4$ is special because it is only
in this dimension that we have $G^{abcd}T_{abcd}=0$ for arbitrary
electromagnetic fields. We may now substitute the expression for
$T$ into the conservation equation above. The coupling of the area
metric background and the gauge theory is then only consistent if
this equation is satisfied. This is in fact to be expected since
the energy momentum tensor was derived from a diffeomorphism
invariant action. In order to see explicitly that this is indeed
the case, we rewrite the resulting conservation condition for
electrodynamics in the form
\begin{equation}
\frac{3}{4}G^{abcd}F_{ab}\nabla^{TM}_{[i}F_{cd]}+\frac{1}{2}F_{di}\left(\nabla^{TM}_c+\frac{1}{2(d-1)}X_c\right)\left(G^{cdab}F_{ab}\right) = 0\,.
\end{equation}
All covariant derivatives in this expression may be replaced with
partial derivatives by using the definition of the dual
electromagnetic induction $(d-2)$-form, see (\ref{induction}). In
this way we obtain
\begin{equation}
\frac{3}{4}G^{abcd}F_{ab}\partial_{[i}F_{cd]}+\frac{1}{2}(-1)^d F_{i_d i}\omega_G^{i_1\dots i_d}\partial_{[i_1}H_{i_2\dots i_{d-1}]}=0\,.
\end{equation}
So the equations of motion of area metric electrodynamics, namely the Bianchi identity $dF=0$, and $dH=0$ imply energy conservation, rendering the matter coupling consistent.

\section{Almost metric manifolds}\label{solar}
A question of immediate interest is of course how standard
Einstein gravity fits into the more general area metric framework,
given its huge phenomenological success especially regarding
observations within our solar system.

More precisely, we would like to know under which circumstances
there exist solutions of~(\ref{generalform}) where the inverse
area metric is of almost metric form $G^{-1} = (G_g^\phi)^{-1}$,
see~(\ref{stablycausallift}), which is the closest one may get to the
standard Lorentzian case. (Recall from section \ref{sec_effmetric}
that the field~$\phi$ is needed in order to render the area metric
induced from a Lorentzian metric normal.)
Interestingly, area metrics with cosmological symmetries are of
almost metric form, as we found in section \ref{sec_Killing}. Thus
the results of the present section will be useful in our
discussion of area metric cosmology in the following
sections~\ref{pfluids}--\ref{sec_exact}. The main aim here, however, is to
study the conditions under which area metric gravity reduces to
Einstein gravity.

In doing so, it is important to withstand the temptation to
discuss the reduction of (\ref{totalaction}) to the almost metric
case at the level of the action. This cannot be meaningful a
priori, because variation with respect to $G$ sweeps out more
variations than variations with respect to a metric and a scalar
field can do \cite{Palais}. If one simply inserted the ansatz
$G^{-1}=(G_g^\phi)^{-1}$ into the total action, this would reduce
the gravitational part to its standard metric analogue, up to a
$\phi$-dependent conformal factor. But we will now show that this
does not correspond to what generically happens at the level of
the full equations of motion including matter.

So we insert the almost metric form $G^{-1}=(G_g^\phi)^{-1}$ into
the full equations of motion, which, as shown in technical detail
in appendix \ref{Qsimplification}, neatly factorizes the cyclic
contribution~$K^C$ of the gravitational variation as
\begin{eqnarray}\label{factorized}
  K^C_{abcd} = S_{[a[c}g_{d]b]} + S_{[c[a}g_{b]d]}\,,
\end{eqnarray}
for some symmetric tensor $S$. If the cyclic contribution $T^C$ of
the energy-momentum tensor to the equations of motion is generated
in like fashion from some symmetric second rank tensor, then the
field equations are equivalent to
\begin{subequations}\label{vacuumGg}
\begin{eqnarray}
\kappa T_{ab} & = &
R_{ab}-\frac{1}{2}Rg_{ab}-\tilde\phi^{-1}\left(\nabla_a\partial_b\tilde{\phi}-g_{ab}\square\tilde{\phi}\right)=\,4K_{ab}\,,\label{vacuumGg1}\\
\kappa T^\phi & = &
-\tilde\phi(1-\tilde\phi^2)^{1/2}R\,,\label{vacuumGg2}
\end{eqnarray}
\end{subequations}
where we define $\tilde{\phi}=(1+\phi^2)^{-1/2}$,
$T_{ab}=4T^C{}^m{}_{amb}$ and $K_{ab}\equiv K^C{}^m{}_{amb}$. When
we discuss cosmological solutions below, we will have to make sure
that $\tilde\phi$ does not leave its allowed range
\begin{equation}\label{constraint}
0\leq\tilde\phi\leq 1\,.
\end{equation}
We will see that this can be understood as a consistency constraint on the initial conditions. It is thus clear that for there to be an almost metric solution at all, the cyclic part of the energy-momentum tensor at hand needs
to factorize in the same fashion as (\ref{factorized}).

In vacuo, this is of course trivially the case. It is also
reassuring that the vacuum system with $T=0$ is causally
well-behaved: in order to see this, we will now perform suitable
field redefinitions, and thus reveal that this system is
conformally equivalent to Einstein-Hilbert gravity minimally
coupled to a massless scalar field. First, using the trace of the
first equation we may replace the system by a simpler one:
\begin{subequations}
\begin{eqnarray}
R_{ab} & = & \tilde\phi^{-1}\nabla_a\partial_b\tilde\phi\,,\\
\square\tilde{\phi} & = & 0\,.
\end{eqnarray}
\end{subequations}
Now a conformal transformation of the metric and a simple
redefinition of the scalar field according to
\begin{equation}
g_{ab}=\tilde{\phi}^{-1/2}\tilde g_{ab}\qquad\textrm{and}\qquad
\Phi=\ln\tilde{\phi}
\end{equation}
shows that the vacuum equations of motion for an area metric of
the almost metric form $G^{-1}=(G_g^\phi)^{-1}$ are conformally
equivalent to the system
\begin{equation}
\tilde R_{ab}=\frac{1}{2}\partial_a\Phi\partial_b\Phi\,,\qquad\square\Phi=0\,.
\end{equation}
These are precisely Einstein's equations for $\tilde g$ coupled to
a massless scalar field $\Phi$, and thus we know that also the
original, conformally related theory is causal. We emphasize again
that the above equations are only valid for almost metric area
metrics in vacuo. This very special case, however, allows for a
most important conclusion: any vacuum solution $(M,g)$ of
Einstein-Hilbert gravity is a vacuum solution of area metric
gravity (setting either $\tilde\phi=1$, or~$\tilde\phi\rightarrow 0$ with
appropriate conditions on the derivatives, see (\ref{vacuumGg})).

Upon the inclusion of matter however, it is no longer true that
metric solutions of Einstein-Hilbert gravity lift to solutions of
area metric gravity. For instance, the area metric energy-momentum
tensor (\ref{edT}) for electrodynamics does not take the
factorized form (\ref{factorized}) for the almost metric ansatz
$G^{-1}=(G_g^\phi)^{-1}$, simply because the first term does not
even contain a metric. Thus, not only can electrodynamics live on a
non-metric area metric spacetime, but also in general
electrodynamics backreacts in such a way as to generate a
non-metric area metric spacetime! This can only be avoided by
restricting the admissible geometries (and thus the variation) to
a standard metric one. In other words, while standard
Einstein-Hilbert solutions are of course stationary points of the
Einstein-Hilbert action, they may fail to be stationary within the
wider spectrum of area metric manifolds, in the presence of matter
whose energy momentum tensor does not factorize.

Thus it is the fact that we vary with respect to the area metric
which causes the potential departure from Einstein-Hilbert gravity
in the presence of matter, while the vacuum solutions of standard
general relativity are also vacuum solutions of area metric
gravity.

\section{Fluid energy momentum}\label{pfluids}
In general relativity perfect fluids present an effective device to discuss the gravitational effects of fundamental matter averaged over large scales. Especially in order to obtain calculationally manageable cosmological models, the description of matter as an ubiquitous fluid is necessary. In this section we will introduce fluid matter on area metric spacetimes as the most general form of energy momentum consistent with area metric cosmology. This discussion paves our way towards a first comparison between area metric cosmology and Einstein cosmology.

The general form of the energy momentum four-tensor is restricted by the Killing symmetries of a given area metric spacetime $(M,G)$. Consider a diffeomorphism of $M$ generated by a vector field $\xi$, and the resulting change $\delta K$ of the variation of the gravitational action with respect to $G^{-1}$. There are two ways to express this quantity. The first simply is the Lie derivative of the four-tensor $K$ along $\xi$, i.e., $\delta K=\mathcal{L}_\xi K$. The second uses the fact that $K$ is completely determined by the area metric $G$; schematically it follows that $\delta K=\delta K/\delta G^{-1}\cdot \mathcal{L}_\xi G^{-1}$. Hence if $X$ is a Killing symmetry of $G$, i.e., if $\mathcal{L}_XG=0$, then we conclude
\begin{equation}
\mathcal{L}_XK_{abcd}=\frac{\delta K_{abcd}}{\delta G^{ijkl}}\left(\mathcal{L}_XG\right)^{ijkl}=0\,.
\end{equation}
The general form of the area metric gravity equations of motion, including matter, is
\begin{equation}
K_{abcd}=\kappa T_{abcd}\,.
\end{equation}
The Lie derivative of the left hand side along any Killing symmetry $X$ of the area metric $G$ vanishes, hence consistency requires that the energy momentum tensor should also satisfy $\mathcal{L}_XT=0$ for each of the background's Killing symmetries. In other words, consistent energy momentum tensors inherit the symmetries of the background. This conclusion is analogous to the one used in standard general relativity to restrict the class of admitted sources in a spacetime with specified symmetry properties.

We may repeat our discussion of section \ref{sec_Killing} to impose the symmetries of an FLRW area metric cosmology on the four-tensor $T$. Recall that $G^{-1}$ is defined in terms of the standard FLRW metric $g$ and a time-dependent scalar $\phi$, and that we have a distinguished time variable $x^a=(x^0,x^\alpha)$. One then finds the components of $T$ to be proportional to those of the area metric up to time-dependent functions. We choose the following parametrization,
\begin{subequations}\label{fluidenergy}
\begin{eqnarray}
T_{0\beta 0\delta} &=& -\frac{F}{1+\phi^2}C_{g\,0\beta 0\delta}\,,\\
T_{0\beta\gamma\delta} &=& \frac{\phi J}{1+\phi^2}\omega_{g\,0\beta\gamma\delta}\,,\\
T_{\alpha\beta\gamma\delta} &=& \frac{N}{1+\phi^2}C_{g\,\alpha\beta\gamma\delta}\,,
\end{eqnarray}
\end{subequations}
where the three functions $F,J$ and $N$ describe the local macroscopic properties of the fluid. Below, we will match these functions with the usual notions of density and pressure for a fluid. But first note that like in the metric case, the fluid energy momentum tensor is not obtained from some Lagrangian, and so the conservation law (\ref{conservation}) is not automatically ensured. We have to impose the energy-momentum conservation condition on the tensor $T$, which, using the Hubble function $H=\dot a/a$, leads to the equation of motion of the fluid,
\begin{equation}\label{fleq}
0=\dot F+\dot J\phi^2-\frac{3\phi\dot\phi\left(F-J\right)}{1+\phi^2}+2H\left(F+N\right).
\end{equation}

Before we turn to our main application of fluid matter, which will be to area metric cosmology, we will introduce in the next section the notion of string fluids which allows us to give a covariant interpretation to fluid matter on area metric spacetimes.

\section{String fluids -- geometry and cosmology}\label{sec_sfluids}
Area metric spacetimes $(M,G)$ do not provide a measure of length which would be required to formulate dynamics for the worldlines of point particles. But they do provide a measure of area, and so one may regard string worldsheets as the minimal dynamical, and hence fundamental, objects on area metric manifolds \cite{Schuller:2005yt}. It is therefore reasonable to expect that any collection of matter fields averaged over large scales should have an approximate description in terms of a fluid made out of strings; this is in complete analogy to standard general relativity where perfect (point particle) fluids are used for this purpose.

Our aim in this section is twofold: first we will formulate a simple energy momentum four-tensor for a string fluid on an arbitrary area metric spacetime, and give a geometric interpretation for the consistency conditions arising from the energy conservation equation~(\ref{conservation}). Second, we will rewrite the general energy momentum tensor consistent with area metric cosmology covariantly in terms of a string fluid; any collection of averaged matter fields in cosmology can thus be understood as a continuous distribution of string worldsheets.

We recall a few basic facts about classical strings on $(M,G)$ from \cite{Schuller:2005ru}. String worldsheets are tangent surfaces to an integrable distribution of areas $\Omega=u\wedge v$ in $A^2TM$. Thus they satisfy the simplicity condtion~(\ref{simple}) and the Frobenius integrability criterion: the commutator of $u$ and $v$ lies within the plane spanned by $\left<u,v\right>$. In components this is equivalent to
\begin{equation}\label{integrability}
\Omega^{p[i}\partial_p\Omega^{jk]}=0\,.
\end{equation}
String dynamics follow from the stationarity of the worldsheet area integral, i.e., from $\int d^2\sigma\sqrt{G(\Omega,\Omega)}=\int d^2\sigma\sqrt{G^C(\Omega,\Omega)}$. The equality follows from the simplicity of $\Omega$, if we explicitly decompose the area metric as
\begin{equation}\label{Gdec}
G_{abcd}=G^C_{abcd}+G^4_{abcd}
\end{equation}
into the cyclic algebraic curvature tensor $G^C$ and a four form $G^4$ in analogy to (\ref{C+F}). It has been shown in \cite{Schuller:2005ru} that the string equation of motion takes a very simple form after fixing the worldsheet reparametrization invariance to constant squared area $G(\Omega,\Omega)$. The stationarity, or minimal surface, condition then simply becomes
\begin{equation}\label{stationarity}
\mathcal{D}_\Omega\Omega (Z) = d[G^C(\Omega,\cdot)](\Omega\wedge Z)=0
\end{equation}
for any vector $Z$. Only the cyclic part of the area metric contributes to this equation.

The energy momentum four-tensor of a string fluid should be expressible in terms of the area metric and the area distribution $\Omega$. We now write down the simplest two terms that obey the required symmetries of an area metric. One of these will then be shown to represent fluids composed of non-interacting strings; the other term is an example for a simple interaction. Define
\begin{equation}\label{sfluid}
T_{abcd}=\frac{1}{4}(\tilde\rho+\tilde p) G_{abij}\Omega^{ij}G_{cdkl}\Omega^{kl}+\frac{1}{3}\tilde p G_{abcd}
\end{equation}
in terms of two functions $\tilde\rho$ and $\tilde p$. In order to
demonstrate the geometric meaning of these terms we analyse the
energy momentum conservation condition. As argued in the preceding
section this condition is not automatically satisfied, but
nevertheless has to be imposed for consistent matter coupling. We
substitute (\ref{sfluid}) into (\ref{conservation}); the resulting
equation can be rewritten as follows:
\begin{equation}
0 = -6(\tilde\rho+\tilde p)\Omega^{cd}\nabla^{LC}_{[i}\left(G_{cd]ab}\Omega^{ab}\right) -4 G_{diab}\Omega^{ab}\left(\nabla^{LC}_c+\frac{1}{6}X_c\right)\left((\tilde\rho+\tilde p) \Omega^{cd}\right)+8\partial_i\tilde p\\
\end{equation}
Observe that the totally antisymmetric part of the area metric in
the first term does not contribute. This is a consequence of the
simplicity (\ref{simple}) and the integrability
(\ref{integrability}) of $\Omega$. Hence we may replace $G$ in the
first term by its cyclic part $G^C$. Due to the antisymmetrization
in this term we may also replace the covariant derivatives by
partials.

Let us discuss the case $\tilde p=0$ first. The first term then
is recognizable as the stationarity condition
(\ref{stationarity}): the local string worlsheets in the string
fluid are stationary, as is classical string motion. The vanishing
of the remaining second term in the string fluid conservation
equation is implied by a generalized continuity equation
\begin{equation}
0 =
\left(\nabla^{LC}_c+\frac{1}{6}X_c\right)\left(\tilde\rho\Omega^{cd}\right)
=
\partial_c\left(|\mathrm{Det}\,G|^{1/6}\tilde\rho\Omega^{cd}\right),
\end{equation}
where the right hand side represents the divergence of a
densitized current. In form-language the continuity equation is
equivalent to $d[\tilde{\rho}\omega_G(\cdot,\Omega)]=0$ for the
two-form $\omega_G(\cdot,\Omega)$. The geometrical picture for
this equation is as follows. Choose any compact four-volume~$V$
with boundary $\partial V$, and any hypersurface defined by
$\Psi=\mathrm{const.}$ with gradient $d\Psi$. The projection on
the hypersurface of the flow of the string fluid through $\partial
V$ is
\begin{equation}
\oint_{\partial V} \tilde{\rho}\omega_G(\cdot,\Omega)\wedge d\Psi
= \int_V d[\tilde{\rho}\omega_G(\cdot,\Omega)]\wedge d\Psi\,.
\end{equation}
This vanishes if the volume $V$ lies entirely within the domain of
the string fluid; then its inflow and outflow through the boundary
$\partial V$ are precisely equal, independent of the choice of
projection on $d\Psi$.

Thus the string fluid energy momentum tensor (\ref{sfluid}) is
amenable to a simple analysis for~$\tilde p=0$: energy momentum
conservation is valid provided that the string fluid obeys both
the stationarity equation and a generalized current conservation,
or continuity, equation. While the worldsheet minimal surfaces of
this situation correspond to zero mean curvature form
$\mathcal{D}_\Omega\Omega=0$, switching on a non-constant pressure
$d\tilde p\neq 0$ means prescribing the local mean curvature by
the projection of the gradient field $d\tilde p$ into the surfaces
$\Omega$. Non-zero~$\tilde p$ (even a constant one) also deforms
the continuity equation. The term proportional to~$\tilde p$ hence
represents a certain class of interactions between the strings
that form the string fluid. Finally, we remark that constant
$\tilde{p}\neq 0$ (with $\tilde\rho=0$) generates
precisely the same terms that one would obtain by adding an area
metric cosmological constant to the gravitational action
$S_{grav}$, in perfect analogy with the metric case.

We now approach our second aim, and part, of this section : the
covariant rewriting of the most general energy momentum tensor
consistent with area metric cosmology, see the preceding section,
in terms of a string fluid.

Recall that area metric cosmologies are of the almost metric type,
featuring the FLRW metric $g$ and an additional time-dependent
scalar $\phi$. String fluid matter in cosmology is tightly
constrained; consistency with isotropy forbids a one-component
string fluid, since the local string worldsheets then would
distinguish a preferred spatial direction. We therefore must
resort to a string fluid with three components
$\Omega_I=\partial_t\wedge v_I$, with three $g$-spacelike
vectors~$v_I$ that may isotropically fill out the spatial sections
in a four-dimensional cosmology. Due to the distinguished
cosmological time variable, $x^a=(x^0,x^\alpha)$, the fluid energy
momentum tensor (\ref{fluidenergy}) has components proportional to
the $0\beta 0\delta$, $0\beta\gamma\delta$ and
$\alpha\beta\gamma\delta$ projections of the area metric,
respectively. These can be defined covariantly using the
three-component string fluid.

Using the decomposition (\ref{Gdec}) we first define a projection (which is an endomorphism on~$\Lambda^2TM$) to the purely spatial indices~$[\alpha\beta]$ by writing
\begin{equation}
2\delta^{[ab]}_{cd}+Q^{ab}{}_{cd}\,,\qquad Q^{ab}{}_{cd}=\sum_I \Omega^{ab}_I G^C_{cdi_1i_2}\Omega^{i_1i_2}_I\,.
\end{equation}
To guarantee the right-action of this operator is a projection we need to satisfy $Q^2=-Q$, which is easily achieved by the requirement
\begin{equation}
G(\Omega_I,\Omega_J)=-\delta_{IJ}\;\Leftrightarrow\;g(\frac{v_I}{\sqrt{1+\phi^2}},\frac{v_j}{\sqrt{1+\phi^2}})=\delta_{IJ}
\end{equation}
which relates the three components of the string fluid. For later convenience we solve this by considering $v^0_I=0$ and $v^\alpha_I/\sqrt{1+\phi^2}$ vielbeins of the three-dimensional metric $g_{\alpha\beta}$: then we have the useful relation $\sum_Iv^\alpha_Iv^\beta_I/(1+\phi^2)=g^{\alpha\beta}$. The second projection we need is the one to the mixed indices $[0\beta]$. It is simply given by the right-action of $-Q$ for which of course $(-Q)^2=-Q$. There is one further way to covariantize the purely spatial components~$G_{g\,\alpha\beta\gamma\delta}$ of the area metric. Consider the operator
\begin{equation}
\tilde Q^{ab}{}_{cd}=\sum_I \Omega^{ab}_I G^4_{cdi_1i_2}\Omega^{i_1i_2}_I\,.
\end{equation}
Since the $\Omega_I$ all contain the preferred time vector $\partial_t$, we have $\Omega_I\wedge \Omega_J=0$, and hence $\tilde Q^2=0$. So $\tilde Q$ is not a projection; it rather maps tensors with mixed indices $[0\beta]$ to tensors with purely spatial indices.

In terms of the three-component string fluid we hence rewrite the fluid energy momentum as a sum of projections of the area metric to its different components, weighted each by time-dependent functions:
\begin{eqnarray}
\!\!\!\!\!\!T_{abcd} & = & G_{p_1p_2r_1r_2}\Big[-\tilde\rho\, Q^{p_1p_2}{}_{ab}Q^{r_1r_2}{}_{cd}+\tilde p \left(2\delta^{p_1p_2}_{ab}+Q^{p_1p_2}{}_{ab}\right)\left(2\delta^{r_1r_2}_{cd}+Q^{r_1r_2}{}_{cd}\right)\Big.\nonumber\\
& &\Big.-\tilde q \left(2\delta^{p_1p_2}_{ab}+Q^{p_1p_2}{}_{ab}\right)Q^{r_1r_2}{}_{cd}-\tilde q\, Q^{p_1p_2}{}_{ab}\left(2\delta^{r_1r_2}_{cd}+Q^{r_1r_2}{}_{cd}\right)+\tilde s\, \tilde Q^{p_1p_2}{}_{ab}\tilde Q^{r_1r_2}{}_{cd}\Big].
\end{eqnarray}
We do not display two further non-vanishing terms involving $\tilde Q Q$ and $\tilde Q(\delta+Q)$ since they simply result in shifts of the functions $\tilde q$ and $\tilde s$. Making use of the schematic relation $\sum_I [G^4(\cdot,\Omega_I) G^C(\cdot,\Omega_I)+G^C(\cdot,\Omega_I) G^4(\cdot,\Omega_I)]=-G^4(\cdot,\cdot)$ which follows from the fact that we have chosen the $v^\alpha_I$ as vielbeins of $g_{\alpha\beta}$, we may rewrite the energy momentum tensor as
\begin{equation}
T_{abcd}  =  (\tilde\rho+\tilde p)\frac{1}{4}\sum_I G_{abij}\Omega^{ij}_IG_{cdkl}\Omega^{kl}_I-(\tilde\rho+\tilde p+\tilde s)\frac{1}{4}\sum_I G^4_{abij}\Omega^{ij}_IG^4_{cdkl}\Omega^{kl}_I+\tilde p\,G_{abcd}+(\tilde\rho+\tilde q)G^4_{abcd}\,.
\end{equation}
Note that we have four functions $\tilde\rho,\tilde p,\tilde q,\tilde s$ here, although three functions $F,J,N$ as in (\ref{fluidenergy}) should be sufficient. Without loss of generality we can in fact fix $\tilde s=-\tilde\rho-\tilde p$. If we do so, and calculate the components $T_{0\beta 0\delta}$, $T_{0\beta\gamma\delta}$ and $T_{\alpha\beta\gamma\delta}$, comparison to (\ref{fluidenergy}) gives the invertible relation
\begin{equation}\label{invrel}
F=\tilde\rho\,,\qquad J=-\tilde q\,,\qquad N=\tilde\rho\phi^2+\tilde p \left(1+\phi^2\right).
\end{equation}

To conclude, the three-component string fluid tensor with $\tilde s=-\tilde\rho-\tilde p$ is a covariant way to rewrite the general source tensor for all types of matter averaged over large scales in an area metric cosmology as
\begin{equation}\label{gensfluid}
T_{abcd}  =  (\tilde\rho+\tilde p)\frac{1}{4}\sum_I G_{abij}\Omega^{ij}_IG_{cdkl}\Omega^{kl}_I + \tilde p\,G_{abcd} + (\tilde\rho+\tilde q)G^4_{abcd}\,.
\end{equation}
The energy momentum conservation equation follows from (\ref{fleq}) and (\ref{invrel}), using for later convenience the redefinition $1+\phi^2=\tilde\phi^{-2}$, and the Hubble function, as
\begin{equation}\label{constring}
0=-\dot{\tilde q}+\left(\dot{\tilde\rho}+\dot{\tilde q}\right)\tilde\phi^2 + 3\left(\tilde\rho+\tilde q\right)\tilde\phi\dot{\tilde\phi} + 2 H\left(\tilde\rho+\tilde p\right).
\end{equation}

A special case of the general expression (\ref{gensfluid}) is the
string fluid representing the sum of three components of
non-interacting strings, see (\ref{sfluid}), which we will call string dust. From our previous
discussion we read off the characteristic relations
\begin{equation}\label{sdust}
\tilde p=0\,,\quad \tilde q=-\tilde\rho\,.
\end{equation}
We will use string dust in our discussion of the late universe,
where interactions presumably are neglibible.

\section{Area metric cosmology}\label{sec_cosmology}
As a first application of the dynamical area metric manifold
theory developed so far, we derive in this section the
equations of area metric cosmology, with the aim to compare our
theory to Einstein's general relativity.

From our discussion of Killing symmetries we know that
cosmological area metrics are of the almost metric type, fully
characterized by just two fields: the FLRW metric $g$ and a
time-dependent scalar field $\phi$. Thus we can use the simplified
equations of motion~(\ref{vacuumGg}) discussed in section
\ref{solar}. These are of the general form
\begin{equation}\label{eqform}
\kappa T_{ab}=4K_{ab}\qquad\textrm{and}\qquad\kappa T^\phi=K^\phi\,.
\end{equation}
We perform a time space split by writing $x^a=(x^0,x^\alpha)$, where greek indices denote spatial coordinates. The characteristic property of the FLRW metric, which obeys $g_{0\alpha}=0$, also implies $K_{0\alpha}=0$. Employing the definition $K_{\beta\delta}=\tilde Kg_{\beta\delta}$, and similarly for the spatial components of the cosmological Ricci tensor, $R_{\beta\delta}=\tilde Rg_{\beta\delta}$, we explicitly obtain the following expressions for the non-vanishing components $K_{00}$ and $\tilde K$:
\begin{subequations}\label{eqgrav}
\begin{eqnarray}
4K_{00} & = & R_{00}+\frac{1}{2}R+3H\dot{\tilde{\phi}}\tilde{\phi}^{-1}\,,\\
4\tilde K & = & \tilde
R-\frac{1}{2}R-(\ddot{\tilde{\phi}}+2H\dot{\tilde{\phi}})\tilde{\phi}^{-1}\,,
\end{eqnarray}
\end{subequations}
where $H=\dot a/a$ is used. The expression for~$K^\phi$ is given by equation (\ref{phiva}).

We now consider the matter contribution to the gravitational equations of motion. As argued above the most general form of matter in area metric cosmology is given by a string fluid energy momentum tensor of the form (\ref{gensfluid}). From the timelike normalizations $G(\Omega_I,\Omega_I)=-1$ we immediately find the string fluid contribution to the scalar equation which simply reads
\begin{equation}
T^\phi=\bar\omega_g^{abcd}T_{abcd}=\frac{24\phi}{1+\phi^2}\tilde q\,.
\end{equation}
The four-tensor contribution $T^C_{abcd}$ is more involved. Since $C^{-1}=G_{g}$ here, this projection to an algebraic curvature tensor is
\begin{eqnarray}\label{astrings}
T^C_{abcd}& = & T_{abcd}+\frac{1}{24}T^\phi\omega_{g\,abcd}+\frac{1}{24}\phi T^\phi C_{g\,abcd}\nonumber\\
& = & \frac{\tilde\rho+\tilde p}{\left(1+\phi^2\right)^2}\left(\sum_I\Omega_{I\,ab}\Omega_{I\,cd} + \phi^2\sum_I\omega_{g\,abij}\Omega^{ij}_I\omega_{g\,cdkl}\Omega^{cd}_I\right) + \frac{\tilde p+\phi^2\tilde q}{1+\phi^2}C_{g\,abcd}\,,
\end{eqnarray}
where the indices on $\Omega_I$ are lowered with the metric $g$.
Importantly there are no non-vanishing components
$T^C_{0\beta\gamma\delta}$. This is guaranteed by our construction
and again shows consistency with the area metric cosmology
background. We proceed to calculate the non-vanishing components
of $T^C$, which again uses the convenient form of the
$\Omega^{0\alpha}_I$ as vielbeins of $g_{\alpha\beta}$. The
results are
\begin{equation}
T^C_{0\beta 0\delta} = \frac{\tilde\rho-\phi^2\tilde q}{1+\phi^2}g_{\beta\delta}\,,\qquad
T^C_{\alpha\beta\gamma\delta} = \frac{\tilde p+\phi^2(\tilde\rho+\tilde p+\tilde q)}{1+\phi^2}C_{g\,\alpha\beta\gamma\delta}\,.
\end{equation}
It is easy to see that $T^C$ is induced by a symmetric two-tensor
in the same way as is~$K^C$ in~(\ref{factorized}). We now
extract the components of $T_{ab}=4T^C{}^m{}_{amb}$; writing
$T_{\alpha\beta}=\tilde Tg_{\alpha\beta}$ this gives
\begin{equation}\label{eqmatt}
T_{00}=12\frac{\tilde{\rho}-\phi^2\tilde{q}}{1+\phi^2}\,,\qquad
\tilde T=4\frac{-\tilde{\rho}+2\tilde p+\phi^2(2\tilde\rho+2\tilde
p+3\tilde q)}{1+\phi^2}\,.
\end{equation}
 According to (\ref{eqform}), we finally combine our results for
the gravitational contributions (\ref{eqgrav}) and our results for
the matter contributions (\ref{eqmatt}) to the general equations
of area metric cosmology coupled to any type of matter. The latter
is described in a parametrization by a three-component string
fluid. We find the simple $\phi$-equation
\begin{equation}\label{Req}
R+24\kappa\tilde q=0\,.
\end{equation}
Calculating the trace of $4K_{ab}=\kappa T_{ab}$ and
using (\ref{Req}), we deduce a standard scalar field equation for $\tilde\phi$,
\begin{equation}\label{phieq}
\square\tilde\phi=\partial V/\partial\tilde\phi\,,
\end{equation}
with potential
\begin{equation}\label{potphi}
V(\tilde\phi)=4\kappa\left(\tilde\rho+\tilde p+\tilde
q\right)\tilde\phi^2-4\kappa\left(\tilde\rho+\tilde
q\right)\tilde\phi^4\,.
\end{equation}

Finally, using both (\ref{Req}) and (\ref{phieq}) to simplify
(\ref{eqform}), we explicitly obtain the time-time and spatial
components of the Ricci tensor
\begin{subequations}\label{graveqs}
\begin{eqnarray}
R_{00} & = & -3H\dot{\tilde\phi}\tilde\phi^{-1}+12\kappa\left(\tilde\rho+\tilde q\right)\tilde\phi^2\,,\\
\tilde R & = & -H\dot{\tilde\phi}\tilde\phi^{-1}-8\kappa\tilde q+4\kappa\left(\tilde\rho+\tilde q\right)\tilde\phi^2\,.
\end{eqnarray}
\end{subequations}

We may now easily compare our equations to those of Einstein gravity, i.e., to {${R_{ab}-Rg_{ab}/2=8\pi G T_{ab}}$}, coupled to a perfect fluid with standard energy momentum {${T_{ab}=(\rho+p) u_a u_b+g_{ab}}$}. Performing the time space split these equations are
\begin{equation}
R_{00}=8\pi G\left(\frac{1}{2}\rho+\frac{3}{2}p\right)\qquad\mathrm{and}\qquad \tilde R=8\pi G\left(\frac{1}{2}\rho-\frac{1}{2}p\right).
\end{equation}
Thus our string fluid gives rise to a perfect fluid with equation of state parameter
\begin{equation}\label{statepar}
w=\frac{p}{\rho}=\frac{x+y}{3(x-y)}\,,\quad x=-H\dot{\tilde\phi}\tilde\phi^{-1}+4\kappa (\tilde\rho+\tilde q)\tilde\phi^2\,,\quad y=4\kappa\tilde q\,.
\end{equation}

Before we analyze this effective behaviour further we have to perform a final consistency check. Since the string fluid energy momentum four-tensor was not derived, by variation with respect to the inverse area metric, from an action, we have to ensure that the energy momentum conservation equation (\ref{constring}) is satisfied. Only then is the coupling of the string fluid to the gravitational equations consistent. To see why this is true, we start from the time-component of the contracted Bianchi identity $\nabla_a R^a{}_{0}-\dot R/2=0$ which is the standard fluid equation of motion ${\dot\rho+3H(\rho+p)=0}$. In our string fluid variables we thus have
\begin{equation}
0 = \partial_t \left(-3H\dot{\tilde\phi}\tilde\phi^{-1}-12\kappa\tilde q+12\kappa\left(\tilde\rho+\tilde q\right)\tilde\phi^2\right)+3H\left(-4H\dot{\tilde\phi}\tilde\phi^{-1}-8\kappa\tilde q+16\kappa\left(\tilde\rho+\tilde q\right)\tilde\phi^2\right).
\end{equation}
We expand the derivative of the term $H\dot{\tilde\phi}\tilde\phi^{-1}$; then we replace $\dot H$ by using the identity $R_{00}=-3\dot H-3H^2$ which holds for any FLRW cosmology, and $\ddot{\tilde\phi}$ by employing the equation of motion of $\tilde\phi$. Up to a sign, this precisely yields the energy momentum conservation condition (\ref{constring}). Our cosmological three-component string fluid thus is consistent without any further restrictions on the functions $\tilde\rho,\tilde p$ and $\tilde q$.

Let us turn to a closer inspection of the effective equation of state parameter $w$. We obtain the following results which are generic for any string fluid, and thus for any type of matter in area metric cosmology (in string fluid parametrization). In the limit $y/x\rightarrow 0$ we obtain the parameter $w=1/3$ of an effective radiation fluid. Note that this limit is exactly realized by the vacuum which, in particular, obeys $y=0$. Thus the vacuum cosmology in area metric geometry is equivalent to Einstein cosmology filled with a radiation fluid.

In the limit $x/y\rightarrow 0$ we obtain the parameter $w=-1/3$ describing a universe with zero acceleration $\ddot a=0$. Consider now the condition $w<-1/3$ for an accelerating universe; it is satisfied if either $y<x<0$ or $y>x>0$. For values of $y$ close to the simple pole in $w$ at $y=x$, we may obtain any value of $w$, both positive or negative. So string fluids in principle should be able to describe any physical universe. Moreover, the diverging effective equation of state parameter $w\rightarrow -\infty$ close to the pole is a temptation to speculate on a possible explanation for the inflationary phase of the universe. Whether this can be realized, however, requires a careful analysis and solutions of the equations.

In the next section we will explicitly analyze the important case of the late universe, which is characterized by non-interacting matter. We will show that area metric cosmology provides a natural explanation for the observed acceleration of our Universe.

\section{String dust cosmology -- the accelerating universe}\label{sec_exact}
In this section we exactly solve the equations of area metric cosmology filled with non-interacting matter, which we describe by string dust. This is the simplest scenario to which we can apply our theory. Moreover, it should describe the late universe, where matter on average has spread out so much that interactions are no longer important.

The defining relations for string dust have been discussed in (\ref{sdust}). We employ these and collect the complete set of equations of motion. The scalar equation of motion follows from~(\ref{potphi}) as
\begin{equation}
0=\ddot{\tilde\phi}+3H\dot{\tilde\phi}\,.
\end{equation}
We also have the gravitational equations (\ref{graveqs}) which, in the string dust case, read
\begin{subequations}\label{gequ}
\begin{eqnarray}
\dot H+H^2 &=& H\dot{\tilde\phi}\tilde\phi^{-1}\,,\\
2ka^{-2}+\dot H+3H^2 &=& -H\dot{\tilde\phi}\tilde\phi^{-1}+8\kappa\tilde\rho\,.
\end{eqnarray}
\end{subequations}
In the non-vacuum case where $\tilde\rho \neq 0$, an equivalent system is obtained by replacing one of the three equations above by the simpler continuity equation which follows from (\ref{constring}), i.e.,
\begin{equation}
0=\dot{\tilde\rho}+2H\tilde\rho\,
\end{equation}
and we will, in fact, replace the scalar equation of motion.

Dividing the first of the gravitational equations (\ref{gequ}) by
$H$ we deduce $Ha/\tilde\phi=\lambda^{-1}$ for constant $\lambda$.
So $\tilde\phi$ is determined by the scale factor. The continuity
equation is easily integrated, and yields $\tilde\rho$ in terms of
the scale factor and a constant $\zeta$. Explicitly we find
\begin{subequations}
\begin{eqnarray}
\tilde\phi &=& \lambda \dot a\,,\\
\tilde\rho &=& \zeta a^{-2}\,.
\end{eqnarray}
\end{subequations}
The remaining equation we have to solve in order to obtain the
string dust solution follows from the second gravitational
equation in (\ref{gequ}):
\begin{equation}\label{aeq}
0=\ddot a+\frac{\dot a^2}{a}+\frac{k-4\kappa\zeta}{a}\,.
\end{equation}
We are interested in the effective equation of state parameter for
the perfect fluid that is created by area metric cosmology in
addition to Einstein cosmology. Hence we rewrite the defining
relations for the parameters $x$ and $y$ in (\ref{statepar}) using
the above equations. This yields
\begin{equation}
x=\frac{\dot a^2+\xi}{a^2}\,,\qquad y=\frac{\xi-k}{a^2}
\end{equation}
for $\xi=k-4\kappa\zeta$. Hence the definition of the parameter $w$ results in
\begin{equation}\label{parst}
w=\frac{1}{3}-\frac{8\kappa}{3}\frac{\zeta}{\dot a^2+k}\,.
\end{equation}
We conclude that once we have a solution of (\ref{aeq}) for the
scale factor $a$, the whole system of equations is integrated, and
provides $\tilde\rho$, $\tilde\phi$ and $w$.

Equation (\ref{aeq}) is exactly solved by the scale factor
\begin{equation}
a(t)=\left\{\begin{array}{cl}
\sqrt{c\left(t-t_0\right)} & \textrm{for }\xi=0\,,\\
\sqrt{c\xi^{-1}-\xi\left(t-t_0\right)^2} & \textrm{for }\xi\neq 0\,,
\end{array}
\right.
\end{equation}
for integration constants $c$ and $t_0$. The parameter $\xi$ is
defined as above. For the sake of completeness, we will now
discuss three possible cases $\xi=0$, $\xi>0$ and $\xi<0$,
assuming $\zeta>0$ (the discussion for
$\zeta<0$ is very similar but does not contain consistent $k=0$
cosmologies). However, it is evident that the only relevant case
for our model of late time cosmology is $\xi<0$, since $\xi=0$
requires fine-tuning, and $\xi>0$ evolves
towards a contracting universe, against the assumed predominance of
the string dust component.

\begin{figure}[ht]
\includegraphics[width=3in]{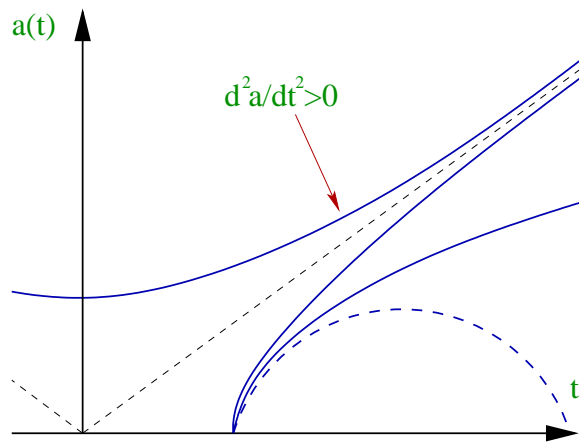}
\caption{\label{figure1}\emph{The solid curves in this sketch show the area metric cosmologies filled with string dust. From bottom to top these are the cases $\xi=0$, $\xi<0$ with $c>0$, and $\xi>0$ with $c<0$. The dashed curve depicts the inconsistent case $\xi>0$. The top curve is an important new result from area geometry (and cannot be obtained as a dust-filled Einstein cosmology): late-time acceleration that tends to zero for $t\rightarrow\infty$.}}
\end{figure}

\paragraph*{(i) Case $\xi=0$.} This case with $k=4\kappa\zeta$ can only be realized by positively curved $k=+1$ cosmologies; this fixes $\zeta$ which is not generic. We also need positive $c>0$ and $t-t_0>0$ for a real solution. We find $\dot a a=c/2$ so that the effective equation of state parameter follows from (\ref{parst}) as
\begin{equation}
w=\frac{1}{3}-\frac{8k\left(t-t_0\right)}{3\left(c+4k\left(t-t_0\right)\right)}\,.
\end{equation}
It is not difficult to show that $w>-1/3$ for all $t>t_0$. In the limit $t\rightarrow\infty$ we find $w\rightarrow -1/3$. Consistency requires $0\leq\tilde\phi\leq 1$, see the discussion for (\ref{constraint}). Since $\tilde\phi=\lambda\dot a$ one can show consistency for $\lambda>0$ and large $t-t_0$. So this case of a string dust filled area metric cosmology describes an open, eternally decelerating, universe with initial singularity, for which the acceleration tends to zero for late times, see figure \ref{figure1}.

\paragraph*{(ii) Case $\xi>0$.} The solution for the scale factor holds for positive $c>0$ and is the upper half of an ellipsoid, defined for $(t-t_0)^2<c/\xi^2$. We deduce $\dot a=-\xi (t-t_0)/a$, but this leads to an inconsistency. For late times, i.e., for $t-t_0\rightarrow \sqrt{c/\xi^2}$, the derivative $\dot a$ diverges, and hence $\tilde\phi$ so that $0\leq\tilde\phi\leq 1$ cannot be satisfied. Therefore this closed universe is not a valid solution.

\paragraph*{(iii) Case $\xi<0$.} This case will turn out to be the most interesting. It has $k<4\kappa\zeta$, and so it can be realized by negatively curved $k=-1$ and flat $k=0$ cosmologies without further restrictions. It can also be realized by $k=+1$ cosmologies, but this requires a matter density $4\zeta>1/\kappa$. Again $\dot a=-\xi (t-t_0)/a$ so that the effective equation of state parameter takes the form
\begin{equation}\label{parg0}
w=\frac{1}{3}-\frac{8\kappa\zeta\left(c-\xi^2\left(t-t_0\right)^2\right)}{3kc-12\kappa\zeta\xi^2\left(t-t_0\right)^2}\,.
\end{equation}
But here, for $\xi<0$, we have to discuss two subcases depending
on the sign of the integration constant $c$. If $c>0$ then the
solution is defined for $t-t_0>\sqrt{c/\xi^2}$. If $c<0$ then the
solution is defined for all $t$. For both signs of $c$ we obtain
the late time limit of $w\rightarrow -1/3$. We can also check
consistency for late times $t-t_0\rightarrow\infty$. If
$0<\lambda\sqrt{-\xi}<1$ this limit ensures $0\leq\tilde\phi\leq
1$. The solution with $c>0$ corresponds to an open, eternally
decelerating, universe with initial singularity, for which the
acceleration tends to zero for late times. The solution for $c<0$
describes an open, eternally \emph{accelerating}, universe without
singularities that passes a minimal radius $a(t_0)=\sqrt{c\xi}$
and has a late-time acceleration tending to zero, see figure
\ref{figure1}. It is worthwhile to note that the sign of $c$
is, in principle, deducible from evaluation of the Hubble
parameter and of the density $\tilde\rho$ at present time, since
$\mathrm{sign}(c)=\mathrm{sign}(\xi+\dot{a}^2)$. In the case of a
flat Universe, for example, acceleration ($c<0$) is obtained if
$4\kappa\tilde\rho_0>H_0^2$ (note, however, that the precise value of
$\kappa$ has still to be fixed, e.g., from comparison with solar
system experiments.)

We conclude that non-interacting string dust matter in area metric cosmology is equivalent to Einstein cosmology filled with a perfect fluid, but, importantly, it provides a more interesting phenomenology than does an Einstein universe filled with standard pressure-free dust. First, area metric cosmology predicts an open late universe, since the closed cosmology is inconsistent. Second, and most strikingly, area metric cosmology  may naturally explain, for all $k=0$ and $k=\pm 1$ cosmologies, the late-time acceleration of the universe! This explanation neither invokes fundamentally unexplained concepts such as a cosmological constant or other forms of dark energy, nor does it invoke fine-tuning since the acceleration automatically tends to small values at late time; these results become consequences of area geometry applied to the very simplest scenario, cosmology.

\section*{CONCLUSIONS}\label{sec_conclusion}
This paper is based on the hypothesis that spacetime is an area metric manifold, which presents a true generalization of metric geometry in dimensions greater than three. By constructing a rigid extension of general relativity to area metric geometry, which can explain the acceleration of the late universe without additional assumptions, we have shown that this idea can be taken to surprisingly interesting conclusions.

The physically immediately relevant case of a four-dimensional area metric manifold is distinguished, since we have shown that the area metric geometry on three-dimensional submanifolds is equivalent to some metric geometry. This reconciles the idea of an area metric spacetime with the phenomenological fact that we can measure lengths and angles in our individual spatial sections. Paying tribute to the phenomenal success of the metric description of spacetime in general relativity, we have constructed curvature invariants that are downward compatible to their metric counterparts. Crucial for this achievement was the extraction of an effective spacetime metric from the fundamental area metric data.

The availability of an area metric curvature scalar, which reduces to the metric Ricci tensor precisely for a metric-induced area metric, in particular allows us to read the Einstein-Hilbert action as dynamics for an area metric. Remarkably, the equations of motion, derived by variation with respect to the area metric, are of second differential order. This circumvention of Lovelock's theorem \cite{Lovelock:1971yv,Lovelock:1972vz} (which for metric manifolds asserts that standard general relativity is the only geometric gravity theory with second order equations) underlines one remarkable aspect of this modification of general relativity.
Another distinguished feature is that the area metric extension of Einstein gravity is rigid, in the sense that it does not use an undetermined deformation (length) scale to add derivative or curvature corrections to the Einstein-Hilbert action, see e.g. \cite{Vollick:2003aw,Carroll:2004de,Schuller:2004rn,Schuller:2004nn,Nojiri:2006su,Punzi:2006bv}, nor does it simply add additional fields propagating on a given background metric, as do scalar tensor theories, see e.g. \cite{Bartolo:1999sq,Coley:1999yq,Esposito-Farese:2000ij,Graf:2006mm}.

We saw that the theory is downward compatible to general relativity in vacuo: every vacuum solution of Einstein's equations also solves the vacuum equations of area metric gravity. Upon the inclusion of matter, however, solutions of Einstein-Hilbert gravity no longer induce solutions of area metric gravity. A case in point is the area metric Einstein-Maxwell theory, which is only consistent for true area metric backgrounds. In other words, not only can electromagnetic fields propagate on an area metric background, but their backreaction requires a deviation from a purely metric spacetime.

The simplest departure from standard metric theory occurs for area metrics of almost metric type, which are determined by a metric and a single scalar. We have shown that the vacuum equations of area metric gravity for this situation are causally well-behaved; conformal transformation and field redefinition demonstrates their equivalence to Einstein gravity coupled to a free scalar field. It should be noted, however, that the equations arising from area metric gravity have to be interpreted without these transformations, since they are not available in the generic, not almost metric, case. The scalar field in the almost metric setting therefore must not be regarded as a field on a metric background, but is itself part of the area metric geometry. The consequences of this mechanism become most apparent in cosmology, which is of the almost metric type, where the scalar in fact behaves as an effective radiation fluid, unlike a standard scalar field.

An area metric structure of spacetime implies that strings, not point particles, are the minimal mechanical objects one may discuss \cite{Schuller:2005yt}. An important consequence of this fact is that the notion of a fluid in cosmology has to be built on an integrable distribution of local worldsheets, rather than on a velocity field for worldlines. We have discussed string fluids arising in this way, and shown that they covariantly describe any collection of matter fields consistent with area metric cosmology. So string fluids can be seen as matter averaged over large scales. We find a remarkable correspondence between, on the one hand, area metric cosmology coupled to the most general consistent string fluid, and, on the other hand, Einstein cosmology coupled to a standard perfect fluid. The energy density and pressure of this effective perfect fluid are given in terms of four functions: the scalar and three parameters of the string fluid. This implies an enormous freedom in the effective equation of state parameter which may, in principle, take any real value.

For the late universe one may assume that matter has spread out so much that interactions are no longer important. We have derived exact solutions for area metric cosmology filled with string dust, i.e., with non-interacting string fluids. While they show some similarity to the solutions of a dust-filled Einstein universe, a remarkable feature emerges that cannot be realized in general relativity: independent of the spatial curvature, area metric cosmology explains the observed small late-time acceleration of our Universe \cite{Spergel:2003cb,Knop:2003iy}. This effect is a consequence of area geometry; it does not invoke a cosmological constant nor additional quintessence fields, see e.g. \cite{Carroll:1998zi,Zlatev:1998tr,Hellerman:2001yi}. Neither does this result depend on fine-tuning, since one of the nice features is that the acceleration automatically tends to zero with time.

So the idea of spacetime as an area metric manifold can be taken very far indeed, has a common conceptual basis in the structure of string and gauge theory, and provides a very encouraging first cosmological prediction, too. But, as can be expected from any novel theoretical framework, a number of open ends remains to be addressed.

The most pressing one from a phenomenological point of view is whether area metric gravity passes solar system tests. This question is certainly solvable, but complicated by the fact that a spherically symmetric area metric is generically not almost metric, but contains one further scalar degree of freedom. While we have shown that every vacuum solution of Einstein gravity, i.e., in particular the Schwarzschild solution, induces a vacuum solution of area metric gravity, it is not clear whether this vacuum solution consistently extends to a solution with matter---having learnt that the inclusion of matter may require deviations from metric geometry. In the metric case life is simpler because of Birkhoff's theorem. Another such physical issue is the discussion of fermions, which might be introduced by their Dirac algebra with respect to the effective metric.

On the mathematical side, a precise understanding of the degrees of freedom united in the area metric is needed. This includes in particular the connection to string theory, where true area metrics emerge as combinations of a metric and a two-form potential.
This problem is likely related to the question of possible relations between area metric geometry and generalized geometries emerging in string theory, such as generalized complex geometry~\cite{Hitchin:2004ut,Gualtieri:2004} or T-folds~\cite{Hull:2004in,Dabholkar:2005ve}. If one ventures to
consider an area metric spacetime structure as a geometrization of the string idea, which we feel is structurally quite natural, one ultimately has to find out how to quantize strings on an area metric background;
given that the trace anomaly in quantized string theory is determined essentially by the two-dimensional worldsheet curvature \cite{Polyakov:1981rd} and that the area metric curvature vanishes in two-dimensions, one may for instance speculate that a successful string quantization based on area metric geometry modifies the standard no-ghost theorems \cite{Polyakov:1981rd,Goddard:1973qh,Polyakov:1981re}. A detailed investigation of these issues is of course needed, and remains the subject of future work.

\acknowledgments{The authors thank Gary Gibbons and Friedrich Hehl
for invaluable discussions; FPS also thanks them for their
generous hospitality, respectively at the Universities of
Cambridge and Cologne. The authors have the pleasure to
acknowlegde useful comments by Robert Beig, Albert Schwarz,
Alexander Turbiner, Julius Wess and extensive discussions with
Marcus Werner and Daniel Sudarsky. RP thanks the Instituto de
Ciencias Nucleares (UNAM) for warm hospitality during the
early stages of this work, and acknowledges financial support from the European Union,
through the `Dottorato di Ricerca' program at the University of
Salerno. The work of FPS has been financially
supported by grant DGAPA-UNAM IN121306. MNRW gratefully
acknowledges full financial support from the German Research
Foundation through the Emmy Noether fellowship grant WO 1447/1-1.}

\appendix
\section{Conventions}\label{conventions}
In this appendix we lay down our conventions for the coordinate representations of objects and operations. Let $M$ be a $d$-dimensional smooth manifold and $\{e_a\}$ a basis of $TM$. This basis canonically induces the basis
\begin{equation}\label{basis1}
\{e_{a_1}\wedge\dots\wedge e_{a_k}\,|\,a_1<\dots<a_k\}
\end{equation}
of $\bigwedge^kTM$ which is of dimension $\left({}^d_k\right)$. Any section $\Omega$ of $\bigwedge^kTM$ can be expanded in this basis in the following form,
\begin{equation}
\Omega=\Omega^{a_1\dots a_k}e_{a_1}\wedge\dots\wedge e_{a_k}=\frac{1}{k!}\Omega^{ab\dots}e_a\wedge e_b\wedge\dots\,,
\end{equation}
which defines the components of $\Omega$. Note that we adhere to the convention that sums over numbered indices $a_1\dots a_k$, \emph{only} on totally \emph{anti}-symmetric objects, are ordered sums over  $a_1<\dots <a_k$. As an immediate consequence of this convention we find the components of exterior products of sections $\Omega\in \bigwedge^kTM$ and $\Sigma\in\bigwedge^lTM$:
\begin{equation}
(\Omega\wedge\Sigma)^{ab\dots}=\left({}^{k+l}_{\;\;k}\right)\Omega^{[ab\dots}\Sigma^{\dots]}\,.
\end{equation}
A basis $\{\epsilon^a\}$ of $T^*M$ induces the canonical basis $\{\epsilon^{a_1}\wedge\dots\wedge\epsilon^{a_k}\,|\,a_1<\dots <a_k\}$ of $\bigwedge^kT^*M$ which is dual to (\ref{basis1}) if $\{\epsilon^a\}$ is dual to $\{e_a\}$, in which case
\begin{equation}
\epsilon^{a_1}\wedge\dots\wedge\epsilon^{a_k}(e_{b_1}\wedge\dots\wedge e_{b_k})=k!\delta^{[a_1}_{[b_1}\dots \delta^{a_k]}_{b_k]}\,,
\end{equation}
where the right hand side are precisely the components of the identity on $\bigwedge^kTM$. We also define contraction symbols for $\Omega\in\bigwedge^{k+1}TM$ and $\omega\in\bigwedge^{l+1}T^*M$:
\begin{subequations}
\begin{eqnarray}
\Omega\,\llcorner\,\omega &=& \textstyle{\frac{1}{k+1}\frac{1}{l+1}}\,\Omega^{a_1\dots a_kp}\omega_{pb_1\dots b_l}e_{a_1}\wedge\dots\wedge e_{a_k}\otimes\epsilon^{b_1}\wedge\dots\wedge\epsilon^{b_k}\,,\\
\omega\,\lrcorner\,\Omega &=& (-1)^{k+l}\,\Omega\,\llcorner\,\omega\,.
\end{eqnarray}
\end{subequations}
For k-forms $\Sigma\in\bigwedge^kT^*M$ we define the exterior derivative such that
\begin{equation}
d\Sigma=(k+1)\partial_{[a_1}\Sigma_{a_2\dots a_{k+1}]}\epsilon^{a_1}\wedge\dots\wedge\epsilon^{a_{k+1}}\,.
\end{equation}

\section{Field equations of area metric gravity}\label{appvary}
This appendix contains full technical detail of the derivation of the equations of motion of area metric gravity in four dimensions. These are obtained from variation, with respect to the inverse area metric, of the action $S_{grav}$ proposed in (\ref{totalaction}) in section \ref{gravtheory}.

This action contains, through the area metric curvature tensor (\ref{areacurv}), covariant derivatives of the $X$-tensors, which we remove by partial integration. For this purpose note that covariant derivatives involving the tangent bundle connection $\nabla^{LC}$ integrated over the area manifold are no longer surface terms, since the area metric is not covariantly constant under this connection. Instead the following formula holds for partial integration:
\begin{equation}\label{partint}
\int_M\omega_G\,\nabla^{LC}_pA^p=-\frac{1}{2(d-1)}\int_M\omega_G\,X_pA^p\,,
\end{equation}
where $X_p=X^{ab}{}_{abp}$. Using this result we obtain the four-dimensional action in the alternative form
\begin{equation}\label{varyac}
\kappa \tilde S_{grav}=\int_M\omega_G\,\left(R(g_G)+\frac{1}{4}g_G^{mn}D^{cdfijk}_{mnabpq}X^{ab}{}_{cdf}X^{pq}{}_{ijk}\right)=\int_M\omega_G\,\tilde L(G)\,,
\end{equation}
which is classically equivalent to the original action on manifolds without boundary. As a shorthand we have used the definition
\begin{equation}\label{Ddef}
D^{cdfijk}_{mnabpq}=\frac{1}{2}\left(\frac{1}{3}\delta^{cdfijk}_{abnpmq}-\frac{1}{3}\delta^{cdfijk}_{abqpmn}+\delta^{cdfijk}_{pmnabq}-\delta^{cdfijk}_{pmqabn}+\left({}^{cdf}_{abm}\leftrightarrow {}^{ijk}_{pqn}\right)\right)
\end{equation}
in contracting the $X$-tensors, where deltas with multiple indices simply denote products of single deltas, e.g. $\delta^{ab}_{cd}=\delta^a_c\delta^b_d$. We proceed to vary the action (\ref{varyac}) with respect to the area metric, which is determined by its decomposition into the irreducible algebraic curvature tensor $C$ and the scalar $\phi$ as in (\ref{scheme}). The calculation is not difficult, but rather long, and will be presented in several steps.

We begin with a variation with respect to the effective metric $g_G$, which of course still depends both on $C$, via the Fresnel tensor, and on $\phi$ via the chosen natural section $d\phi$ of the cotangent bundle. The dependence on the effective metric in all covariant derivatives $\nabla^{LC}$, also in the tensors $X$, has to be taken account. Variation of the Levi-Civita connection and of the corresponding Riemann curvature tensor with respect to a given metric are standard. We explicitly provide these variations with respect to the inverse metric. Then, for connection and curvature
\begin{subequations}\label{h1}
\begin{eqnarray}
\delta_{g_G^{-1}}\Gamma^t_{vk} & = & -\frac{1}{2}g_G^{tu}g_{G\,y\alpha}g_{G\,z\beta}\tilde D^{xyz}_{kvu}\nabla^{LC}_x\delta g_G^{\alpha\beta}\,,\\
\delta_{g_G^{-1}}R_{mn} & = & -\frac{1}{2}g_G^{tu}g_{G\,y\alpha}g_{G\,z\beta}\left(\tilde D^{xyz}_{nmu}\delta^w_t-\tilde D^{xyz}_{tmu}\delta^w_n\right)\nabla^{TM}_w\nabla^{TM}_x\delta g_G^{\alpha\beta}\,,
\end{eqnarray}
\end{subequations}
where we use another notational shorthand,
\begin{equation}
\tilde D^{xyz}_{kvu}=\delta^{xyz}_{kvu}+\delta^{xyz}_{vku}-\delta^{xyz}_{uvk}\,.
\end{equation}
The same remark as above applies for deltas with multiple indices. The variation of the $X$-tensor with respect to the effective metric follows as
\begin{equation}
\delta_{g_G^{-1}}X^{pq}{}_{ijk} =  g_G^{tu}g_{G\,y\alpha}g_{G\,z\beta}\left(\frac{1}{4}\tilde D^{xyz}_{kvu}G^{pqvw}G_{twij}+\tilde D^{xyz}_{k[i|u}\delta^{[pq]}_{t|j]}\right)\nabla^{LC}_x\delta g_G^{\alpha\beta}\,.\label{h3}
\end{equation}
Employing these relations and performing all necessary partial integrations one hence obtains the expression
\begin{eqnarray}
\delta_{g_G^{-1}}(\kappa\tilde S_{grav}) & = & \int_M\omega_G\,\delta g_G^{\alpha\beta}V_{g_G\,\alpha\beta}\,,\nonumber\\
V_{g_G\,\alpha\beta} & = &
R(g_G)_{\alpha\beta}+\frac{1}{4}D^{cdfijk}_{\alpha\beta abpq}X^{ab}{}_{cdf}X^{pq}{}_{ijk} - \frac{1}{2}g_G^{mn}g_G^{tu}g_{G\,y\alpha}g_{G\,z\beta}D^{cdfijk}_{mnabpq}\times \nonumber\\
& & \quad \times\left(\nabla^{LC}_x+\frac{1}{6}X_x\right)\left(X^{ab}{}_{cdf}\left(\frac{1}{4}\tilde D^{xyz}_{kvu}G^{pqvw}G_{twij}+\tilde D^{xyz}_{k[i|u}\delta^{[pq]}_{t|j]} \right)\right) \nonumber\\
& & {} - \frac{1}{6}\left(\delta^{xw}_{\alpha\beta}-g_G^{xw}g_{G\,\alpha\beta}\right)\left(\nabla^{LC}_xX_w+\frac{1}{6}X_xX_w\right).
\end{eqnarray}

Our second step is the variation of the action with respect to all explicitly occurring area metrics; these appear in the volume form, and also in the $X$-tensors. Explicitly we have
\begin{subequations}
\begin{eqnarray}
\delta_{G^{-1}}\omega_G & = & -\frac{\omega_G}{8(d-1)}\,G_{\alpha\beta\gamma\delta}\delta G^{\alpha\beta\gamma\delta}\,,\label{G1}\\
\delta_{G^{-1}}X^{pq}{}_{ijk} & = & -\frac{1}{4}G_{ij\gamma\delta}\left(X^{pq}{}_{\alpha\beta k}\delta G^{\alpha\beta\gamma\delta}+\delta^{pq}_{\alpha\beta}\nabla^{LC}_k\delta G^{\alpha\beta\gamma\delta}\right).\label{G2}
\end{eqnarray}
\end{subequations}
Application of these two identities finally yields the expression
\begin{eqnarray}
\delta_{G^{-1}}(\kappa\tilde S_{grav}) & = & \int_M\omega_G\,\delta G^{\alpha\beta\gamma\delta}V_{G\,\alpha\beta\gamma\delta}\,,\nonumber\\
V_{G\,\alpha\beta\gamma\delta} & = & -\frac{1}{24}G_{\alpha\beta\gamma\delta}\tilde L(G)
-\frac{1}{8}g_G^{mn}D^{cdfijk}_{mnabpq}X^{ab}{}_{cdf}X^{pq}{}_{\alpha\beta k}G_{ij\gamma\delta} \nonumber\\
& & {} +\frac{1}{8}g_G^{mn}D^{cdfijk}_{mnab\alpha\beta}\left(\nabla^{LC}_k+\frac{1}{6}X_k\right)\left(X^{ab}{}_{cdf}G_{ij\gamma\delta}\right).
\end{eqnarray}

In order to obtain the full variation of the action, we last but not least have to consider the dependence of the effective metric and of the area metric itself on the irreducible parts~$C$
and~$\phi$ of the area metric, and then to combine all terms. For the area metric we find
\begin{equation}\label{varG}
\delta G^{ijkl} = \delta C^{ijkl}+\frac{1}{24}\,\phi\,\omega_C^{ijkl}C^{-1}_{\alpha\beta\gamma\delta}\delta C^{\alpha\beta\gamma\delta} + \omega_C^{ijkl}\delta\phi\,.
\end{equation}
The expressions for the effective metric are more elaborate. In our mainly plus signature convention these are given by
\begin{equation}
\delta_\phi g_G^{\alpha\beta}  =  6\left(2h^{-5}\mathcal{G}^\alpha\mathcal{G}^\beta\mathcal{G}^\gamma-3h^{-3}\mathcal{G}^{(\alpha\beta}\mathcal{G}^\gamma)+h^{-1}\mathcal{G}^{\alpha\beta\gamma}\right)\nabla^{LC}_\gamma\delta\phi \label{hphi}
\end{equation}
and
\begin{eqnarray}
\delta_C g_G^{ab} & = & \left(-3h^{-5}(\partial\phi)^4_{ijkl}\mathcal{G}^a\mathcal{G}^b+\frac{3}{2}h^{-3}(\partial\phi)^4_{ijkl}\mathcal{G}^{ab}+4h^{-3}\delta^{(a}_{(i}(\partial\phi)^3_{jkl)}\mathcal{G}^{b)}-3h^{-1}\delta^a_{(i}\delta^b_{j}(\partial\phi)^2_{kl)}\right)\!\times\nonumber\\
& & {}\times \left(\mathcal{G}^{ijkl}C^{-1}_{\alpha\beta\gamma\delta}+\omega_{C\,mnp\beta}\omega_{C\,rs\gamma\delta}C^{mnri}C^{jpsk}\delta^l_\alpha+\frac{1}{2}\omega_{C\,mn\beta q}\omega_{C\,r\gamma tu}C^{mnri}C^{lqtu}\delta^j_\alpha\delta^k_\delta\right)\times\nonumber\\
& & {}\times\left(\frac{1}{12}\,\delta C^{\alpha\beta\gamma\delta}\right),
\label{hC}
\end{eqnarray}
where the characteristic function $h$ evaluated on the section $d\phi$, see (\ref{character}), appears. We also use the abbreviations $\mathcal G^{\alpha\beta\gamma}=\mathcal G^{\alpha\beta\gamma d}\partial_d\phi$, $\mathcal G^{\alpha\beta}=\mathcal G^{\alpha\beta cd}(\partial\phi)^2_{cd}$ and $\mathcal G^\alpha=\mathcal G^{\alpha bcd}(\partial\phi)^3_{bcd}$.

The final result for the combined terms resulting from the total variation of the action $\tilde S_{grav}$ may now be written in the simple form
\begin{equation}\label{result}
\delta (\kappa\tilde S_{grav}) = \int_M\omega_G\,\left(\delta\phi K^\phi+\delta C^{\alpha\beta\gamma\delta}K^C_{\alpha\beta\gamma\delta}\right).
\end{equation}
Note that in reaching this result another partial integration has to be performed on the terms arising from the variation $\delta_\phi g_G$. The final result for $K^\phi$ is then given by the expression
\begin{equation}\label{Vphi}
K^\phi=-\left(\nabla^{LC}_\gamma+\frac{1}{6}X_\gamma\right)\left(6V_{g_G\,\alpha\beta}\left(2h^{-5}\mathcal{G}^\alpha\mathcal{G}^\beta\mathcal{G}^\gamma-3h^{-3}\mathcal{G}^{(\alpha\beta}\mathcal{G}^{\gamma)}+h^{-1}\mathcal{G}^{\alpha\beta\gamma}\right)\right)+\omega_C^{\alpha\beta\gamma\delta}V_{G\,\alpha\beta\gamma\delta}\,.
\end{equation}
The almost final result for $K^C$ is given by the following expression $\tilde K^C$ which, however, does not yet have the correct symmetry structure:
\begin{eqnarray}\label{VCalmost}
\tilde K^C_{\alpha\beta\gamma\delta} & = & \left(-3h^{-5}(\partial\phi)^4_{ijkl}\mathcal{G}^a\mathcal{G}^b+\frac{3}{2}h^{-3}(\partial\phi)^4_{ijkl}\mathcal{G}^{ab}+4h^{-3}\delta^{(a}_{(i}(\partial\phi)^3_{jkl)}\mathcal{G}^{b)}-3h^{-1}\delta^a_{(i}\delta^b_{j}(\partial\phi)^2_{kl)}\right)\!\times\nonumber\\
& & {}\times \left(\mathcal{G}^{ijkl}C^{-1}_{\alpha\beta\gamma\delta}+\omega_{C\,mnp\beta}\omega_{C\,rs\gamma\delta}C^{mnri}C^{jpsk}\delta^l_\alpha+\frac{1}{2}\omega_{C\,mn\beta q}\omega_{C\,r\gamma tu}C^{mnri}C^{lqtu}\delta^j_\alpha\delta^k_\delta\right)\times\nonumber\\
& & {}\times \left(\frac{1}{12}V_{g_G\,ab}\right)
+V_{G\,\alpha\beta\gamma\delta}+\frac{1}{24}\phi V_{G\,ijkl}\omega_C^{ijkl}C^{-1}_{\alpha\beta\gamma\delta}\,.
\end{eqnarray}
To obtain the correct $K^C$ note that this term must have the symmetries of an algebraic curvature tensor since these symmetries are imposed by its contraction with $\delta C$ in equation (\ref{result}). To correct procedure thus is to project onto this irreducible representation by removing the totally antisymmetric part:
\begin{equation}
K^C_{ijkl}= \tilde K^C_{ijkl}+\frac{1}{24}\omega_C^{\alpha\beta\gamma\delta}\tilde K^C_{\alpha\beta\gamma\delta}\omega_{C\,ijkl}\,.
\end{equation}
The positive sign between both terms is a consequence of the Lorentzian signature.

Finally, we arrive at the vacuum gravity equations for four-dimensional area metric manifolds. Vanishing variation $\delta \tilde S_{grav}=0$ with respect to the inverse area metric implies the following two equations, corresponding to its cyclic and totally antisymmetric parts, respectively:
\begin{equation}\label{vac}
K^C_{ijkl}=0\quad\mathrm{and}\quad K^\phi=0\,.
\end{equation}

\section{Simplification for almost metric area metrics}\label{Qsimplification}
In section \ref{solar} we have discussed almost metric area metrics $G^{-1}=(G_g^\phi)^{-1}$, which are determined by a metric and scalar field, as those closest to standard Lorentzian manifolds. For this rather simple case occurs a significant simplification of the four-dimensional equations of motion that have been derived in detail in the preceding section. We will now present this simplification in some more detail as a basis for our discussion of area metric cosmology in the main text.

Essentially because of the simple form (\ref{metricFresnel}) of the Fresnel tensor in the almost metric case, we find that the inner bracket in the $\phi$-variation, see $K^\phi$ in (\ref{Vphi}), is identically zero, and hence its derivative is. Hence
\begin{equation}\label{phisimple}
K^\phi=\bar\omega_g^{ijkl}V_{G\,ijkl}\,.
\end{equation}
In vacuo this expression must vanish by equation (\ref{vac}), which simply requires that $V_G$ should be an algebraic curvature tensor.

A similarly drastic simplification occurs for the variation with respect to $C$. We expand all terms that multiply $V_{g_G\,ab}$ in equation (\ref{VCalmost}), while at the same time imposing the algebraic curvature tensor symmetries on the indices $\alpha\beta\gamma\delta$. This results in
\begin{equation}
-\frac{1}{12}g_G^{ab}g_{G_\alpha[\gamma}g_{G\,\delta]\beta}+\frac{1}{2}\delta^{(ab)}_{\alpha[\gamma}g_{G\,\delta]\beta}\,.
\end{equation}
The simplified form of the $C$-equation then follows as
\begin{eqnarray}
K^C_{\alpha\beta\gamma\delta} & = & -\frac{1}{12}g_G^{ab}V_{g_G\,ab}g_{G\,\alpha[\gamma}g_{G\,\delta]\beta}+\frac{1}{2}V_{g_G\,[\alpha[\gamma}g_{G\,\delta]\beta]}\nonumber\\
& & {}+\frac{1}{12}\phi K^\phi g_{G\,\alpha[\gamma}g_{G\,\delta]\beta}+V_{G\,\alpha\beta\gamma\delta}+\frac{1}{24}K^\phi\omega_{\alpha\beta\gamma\delta}\,.
\end{eqnarray}
The antisymmetrization in the second term is supposed to act only on the index pairs $[\alpha\beta]$ and $[\gamma\delta]$. This equation indeed has the symmetries of an algebraic curvature tensor; this is ensured by the addition of the last term which removes the totally antisymmetric part of~$V_G$.

The inverse $G_g^\phi$ of $(G_g^\phi)^{-1}$ has already been displayed in (\ref{invlift}). This expression immediately enables us to calculate the $X$-tensor as
\begin{equation}
X^{ab}{}_{cdf}=-\frac{\partial_f \phi}{1+\phi^2}\left(\frac{1}{2}\omega^{ab}{}_{cd}+\phi\delta^{[ab]}_{cd}\right).
\end{equation}
We now have all the information needed to further simplify the above expressions for $K^\phi$ and $K^C$ in the almost metric case. For the $\phi$-variation we simply find
\begin{equation}\label{phiva}
K^\phi=-\frac{\phi}{1+\phi^2}R\,.
\end{equation}
The $C$-variation $K^C$ which is a rank four tensor turns out to conform to a very particular structure. It is in fact induced from a symmetric rank two tensor $S$ as shown in equation~(\ref{factorized}), where
\begin{equation}
4S_{ab} = R_{ab}-\frac{1}{3}Rg_{ab}+\frac{1-2\phi^2}{\left(1+\phi^2\right)^2}\left(\partial_a\phi\partial_b\phi-\frac{1}{2}g_{ab}(\partial\phi)^2\right) +\frac{\phi}{1+\phi^2}\left(\nabla_a\partial_b\phi-\frac{1}{2}g_{ab}\square\phi\right).
\end{equation}
Our main application of these expressions in this paper is to area metric cosmology.\\



\thebibliography{00}
\bibitem{Spergel:2003cb}
  D.~N.~Spergel {\it et al.}  [WMAP Collaboration],
  Astrophys.\ J.\ Suppl.\  {\bf 148} (2003) 175
  [arXiv:astro-ph/0302209].

\bibitem{Knop:2003iy}
  R.~A.~Knop {\it et al.}  [Supernova Cosmology Project Collaboration],
  Astrophys.\ J.\  {\bf 598} (2003) 102
  [arXiv:astro-ph/0309368].

\bibitem{Drummond:1979pp}
  I.~T.~Drummond and S.~J.~Hathrell,
  Phys.\ Rev.\ D {\bf 22} (1980) 343.

\bibitem{Schuller:2005ru}
  F.~P.~Schuller and M.~N.~R.~Wohlfarth,
  JHEP {\bf 0602}, 059 (2006)
  [arXiv:hep-th/0511157].

\bibitem{Fradkin:1985qd}
  E.~S.~Fradkin and A.~A.~Tseytlin,
  Phys.\ Lett.\ B {\bf 163} (1985) 123.

\bibitem{Abouelsaood:1986gd}
  A.~Abouelsaood, C.~G.~Callan, C.~R.~Nappi and S.~A.~Yost,
  Nucl.\ Phys.\ B {\bf 280} (1987) 599.

\bibitem{Bergshoeff:1987at}
  E.~Bergshoeff, E.~Sezgin, C.~N.~Pope and P.~K.~Townsend,
  Phys.\ Lett.\  {\bf 188B} (1987) 70.

\bibitem{Leigh:1989jq}
  R.~G.~Leigh,
  Mod.\ Phys.\ Lett.\ A {\bf 4}, 2767 (1989).

\bibitem{Ashtekar:1986yd}
  A.~Ashtekar,
  Phys.\ Rev.\ Lett.\  {\bf 57}, 2244 (1986).

\bibitem{Rovelli:1989za}
  C.~Rovelli and L.~Smolin,
  Nucl.\ Phys.\ B {\bf 331}, 80 (1990).

\bibitem{Rovelli:1994ge}
  C.~Rovelli and L.~Smolin,
  Nucl.\ Phys.\ B {\bf 442}, 593 (1995)
  [Erratum-ibid.\ B {\bf 456}, 753 (1995)]
  [arXiv:gr-qc/9411005].

\bibitem{Schuller:2005yt}
  F.~P.~Schuller and M.~N.~R.~Wohlfarth,
  Nucl.\ Phys.\ B {\bf 747}, 398 (2006)
  [arXiv:hep-th/0508170].

\bibitem{Kachru:2002sk}
  S.~Kachru, M.~B.~Schulz, P.~K.~Tripathy and S.~P.~Trivedi,
  JHEP {\bf 0303} (2003) 061
  [arXiv:hep-th/0211182].

\bibitem{Hellerman:2002ax}
  S.~Hellerman, J.~McGreevy and B.~Williams,
  JHEP {\bf 0401} (2004) 024
  [arXiv:hep-th/0208174].

\bibitem{Gurrieri:2002wz}
  S.~Gurrieri, J.~Louis, A.~Micu and D.~Waldram,
  Nucl.\ Phys.\ B {\bf 654} (2003) 61
  [arXiv:hep-th/0211102].

\bibitem{Fidanza:2003zi}
  S.~Fidanza, R.~Minasian and A.~Tomasiello,
  Commun.\ Math.\ Phys.\  {\bf 254} (2005) 401
  [arXiv:hep-th/0311122].

\bibitem{Dabholkar:2002sy}
  A.~Dabholkar and C.~Hull,
  JHEP {\bf 0309} (2003) 054
  [arXiv:hep-th/0210209].

\bibitem{Hull:2003kr}
  C.~M.~Hull and A.~Catal-Ozer,
  JHEP {\bf 0310} (2003) 034
  [arXiv:hep-th/0308133].

\bibitem{Flournoy:2004vn}
  A.~Flournoy, B.~Wecht and B.~Williams,
  Nucl.\ Phys.\ B {\bf 706} (2005) 127
  [arXiv:hep-th/0404217].

\bibitem{Hitchin:2004ut}
  N.~Hitchin,
  Quart.\ J.\ Math.\ Oxford Ser.\  {\bf 54} (2003) 281
  [arXiv:math.dg/0209099].

\bibitem{Gualtieri:2004}
  M.~Gualtieri,
  {\sl Generalized complex geometry},
  Oxford University DPhil thesis,
  [arXiv:math.dg/0401221].

\bibitem{Hull:2004in}
  C.~M.~Hull,
  JHEP {\bf 0510} (2005) 065
  [arXiv:hep-th/0406102].

\bibitem{Dabholkar:2005ve}
  A.~Dabholkar and C.~Hull,
  JHEP {\bf 0605} (2006) 009
  [arXiv:hep-th/0512005].

\bibitem{Grana:2004bg}
  M.~Grana, R.~Minasian, M.~Petrini and A.~Tomasiello,
  JHEP {\bf 0408} (2004) 046
  [arXiv:hep-th/0406137].

\bibitem{Grana:2005ny}
  M.~Grana, J.~Louis and D.~Waldram,
  arXiv:hep-th/0505264.

\bibitem{Koerber:2005qi}
  P.~Koerber,
  JHEP {\bf 0508} (2005) 099
  [arXiv:hep-th/0506154].

\bibitem{Zucchini:2005rh}
  R.~Zucchini,
  JHEP {\bf 0503} (2005) 022
  [arXiv:hep-th/0501062].

\bibitem{Zabzine:2006uz}
  M.~Zabzine,
  arXiv:hep-th/0605148.

\bibitem{Reid-Edwards:2006vu}
  R.~A.~Reid-Edwards,
  arXiv:hep-th/0610263.
  
\bibitem{Grange:2006es}
  P.~Grange and S.~Schafer-Nameki,
  arXiv:hep-th/0609084.

\bibitem{Becker:2006ks}
  K.~Becker, M.~Becker, C.~Vafa and J.~Walcher,
  arXiv:hep-th/0611001.

\bibitem{Lammerzahl:2004ww}
  C.~Lammerzahl and F.~W.~Hehl,
  Phys.\ Rev.\ D {\bf 70} (2004) 105022
  [arXiv:gr-qc/0409072].

\bibitem{Hehl:2004yk}
  F.~W.~Hehl and Y.~N.~Obukhov,
  Found.\ Phys.\  {\bf 35} (2005) 2007
  [arXiv:physics/0404101].

\bibitem{Hehl:2005xu}
  F.~W.~Hehl, Yu.~N.~Obukhov, G.~F.~Rubilar and M.~Blagojevic,
  Phys.\ Lett.\ A {\bf 347} (2005) 14
  [arXiv:gr-qc/0506042].

\bibitem{Peres:1962}
  A.~Peres,
  Ann.\ Phys.\ {\bf 19} (1962) 279.

\bibitem{foot1}
This continues the canonical approach of \cite{Schuller:2005ru}, whereas an example of additional structure is provided by multi-metric backgrounds interpreted as an area metric \cite{Schuller:2005yt}.

\bibitem{Finslerbook}
  M.~Abate and G.~Patrizio,
  {\sl Finsler metrics---a global approach},
  Lecture\ notes\ in\ mathematics\ {\bf 1591},
  Springer, Berlin 1994.

\bibitem{Cartan:1933}
  E.~Cartan,
  {\sl Les espaces m\'etriques fond\'es sur la notion d'aire},
  Hermann, Paris 1933.

\bibitem{Barthel:1959}
  W.~Barthel,
  Math.\ Annalen\ {\bf 137} (1959) 42.

\bibitem{Brickell:1968}
  F.~Brickell,
  Can.\ J.\ Math.\ {\bf 90} (1968) 540.

\bibitem{Davies:1972}
  E.~T.~Davies,
  Aequ.\ Math.\ {\bf 7} (1972) 173.

\bibitem{Davies:1973}
  E.~T.~Davies,
  Atti\ Accad.\ Naz.\ dei\ Linzei\ {\bf 53} (1973) 389.

\bibitem{Tamassy:1995}
  L.~Tam\'assy,
  Rep.\ Math.\ Phys.\ {\bf 36} (1995) 453.

\bibitem{Isham}
  C.~J.~Isham,
  {\sl Modern differential geometry for physicists},
  World\ Scientific\ Lecture\ Notes\ {\bf 61},
  World Scientific, Singapore 1999.

\bibitem{Beem}
  J.~K.~Beem, P.~Ehrlich and K.~Easley,
  {\sl Global Lorentzian geometry}, 2nd ed.,
  Marcel Dekker, New York 1996.

\bibitem{Lovelock:1971yv}
  D.~Lovelock,
  J.\ Math.\ Phys.\  {\bf 12} (1971) 498.

\bibitem{Lovelock:1972vz}
  D.~Lovelock,
  J.\ Math.\ Phys.\  {\bf 13} (1972) 874.

\bibitem{Palais}
R.~S.~Palais, Comm.\ Math.\
Physics {\bf69} (1979) 19.

\bibitem{Vollick:2003aw}
  D.~N.~Vollick,
  Phys.\ Rev.\ D {\bf 68} (2003) 063510
  [arXiv:astro-ph/0306630].

\bibitem{Carroll:2004de}
  S.~M.~Carroll, A.~De Felice, V.~Duvvuri, D.~A.~Easson, M.~Trodden and M.~S.~Turner,
  Phys.\ Rev.\ D {\bf 71} (2005) 063513
  [arXiv:astro-ph/0410031].

\bibitem{Schuller:2004rn}
  F.~P.~Schuller and M.~N.~R.~Wohlfarth,
  Nucl.\ Phys.\ B {\bf 698} (2004) 319
  [arXiv:hep-th/0403056].

\bibitem{Schuller:2004nn}
  F.~P.~Schuller and M.~N.~R.~Wohlfarth,
  Phys.\ Lett.\ B {\bf 612} (2005) 93
  [arXiv:gr-qc/0411076].

\bibitem{Nojiri:2006su}
  S.~Nojiri and S.~D.~Odintsov,
  arXiv:hep-th/0610164.

\bibitem{Punzi:2006bv}
  R.~Punzi, F.~P.~Schuller and M.~N.~R.~Wohlfarth,
  arXiv:gr-qc/0605017.

\bibitem{Bartolo:1999sq}
  N.~Bartolo and M.~Pietroni,
  Phys.\ Rev.\ D {\bf 61} (2000) 023518
  [arXiv:hep-ph/9908521].

\bibitem{Coley:1999yq}
  A.~A.~Coley,
  Gen.\ Rel.\ Grav.\  {\bf 31} (1999) 1295
  [arXiv:astro-ph/9910395].

\bibitem{Esposito-Farese:2000ij}
  G.~Esposito-Farese and D.~Polarski,
  Phys.\ Rev.\ D {\bf 63} (2001) 063504
  [arXiv:gr-qc/0009034].

\bibitem{Graf:2006mm}
  W.~Graf,
  arXiv:gr-qc/0602054.

\bibitem{Carroll:1998zi}
  S.~M.~Carroll,
  Phys.\ Rev.\ Lett.\  {\bf 81} (1998) 3067
  [arXiv:astro-ph/9806099].

\bibitem{Zlatev:1998tr}
  I.~Zlatev, L.~M.~Wang and P.~J.~Steinhardt,
  Phys.\ Rev.\ Lett.\  {\bf 82} (1999) 896
  [arXiv:astro-ph/9807002].

\bibitem{Hellerman:2001yi}
  S.~Hellerman, N.~Kaloper and L.~Susskind,
  JHEP {\bf 0106} (2001) 003
  [arXiv:hep-th/0104180].

\bibitem{Polyakov:1981rd}
  A.~M.~Polyakov,
  Phys.\ Lett.\ B {\bf 103} (1981) 207.

\bibitem{Goddard:1973qh}
  P.~Goddard, J.~Goldstone, C.~Rebbi and C.~B.~Thorn,
  Nucl.\ Phys.\ B {\bf 56}(1973) 109.

\bibitem{Polyakov:1981re}
  A.~M.~Polyakov,
  Phys.\ Lett.\ B {\bf 103} (1981) 211.
\end{document}